\documentclass{aa}  

\usepackage{graphicx}
\usepackage{txfonts}
\usepackage{xcolor}
\usepackage{todonotes}
\usepackage{dblfloatfix}
\usepackage{float}

\usepackage{tabularx}
\usepackage{booktabs}
\usepackage{array} 
\usepackage[colorlinks=true, allcolors=black]{hyperref}

\usepackage{tikz}

\usepackage[normalem]{ulem} 
\begin{document}

   \title{Statistical Study of Balmer Continuum Enhancement in Solar Flares}

   \author{Pranjali Sharma
          \inst{1}
          \and
          Lucia Kleint
          \inst{1}
          \and
          Jonas Zbinden 
          \inst{1}
                  }

   \institute{Astronomical Institute of the University of Bern, Sidlerstrasse 5, 3012 Bern, Switzerland\\
              \email{Pranjali.sharma@unibe.ch}
                     }


 
\abstract
{Identifying the physical mechanisms of continuum emission in solar flares 
is important to improve our understanding of the transport of energy in the chromosphere.
This requires reliable detection of enhanced continuum emission across different flare classes.}
{This study aims to quantify the occurrence statistics and spatial and temporal characteristics of near-ultraviolet (NUV) continuum enhancements across various classes of solar flares.}
{We analyzed 234 IRIS flare observations and developed two independent detection pipelines. Both pipelines initially extract candidate enhancement events from pixel-level NUV time series and subsequently eliminate false positives by making use of the temporal and spatial correspondence between NUV and FUV continuum enhancement. For one of the pipelines, we used Gaussian process regression to quantify the uncertainty in the enhancement magnitudes.}
{We detected NUV continuum enhancements in 80 out of 234 flares. The enhancements occurred predominantly on the flare ribbon edges and during the GOES impulsive phase but also after the GOES peak flux.
In a few cases (4 pixels), NUV and FUV continuum enhancement was detected 7–15 minutes before the GOES start or more than 20 minutes after the peak, appearing as indistinct bright points in the active regions. Despite large uncertainties for C-class events, enhancement magnitudes increase with flare class, with X-class flares showing the strongest enhancement.}
{Our analysis reveals that the enhancements are confined to localized regions on the flare ribbon edges. In terms of flare energetics, this suggests the possibility of enhancement occurring preferably in the regions with freshly reconnected magnetic field lines, or the ribbon fronts with gradual and modest high-energy flux injection of the non-thermal electrons. Enhancements found significantly after the flare peak in strong flares further suggest multiple heating episodes. The enhancement strengths of flare events as weak as C1.1 from this study serve as an important constraint for flare simulation models. }


\maketitle
\nolinenumbers
%

\section{Introduction}

In stark contrast to the relatively stable photosphere, the chromosphere and corona host highly energetic phenomena, one of which are solar flares. The energy released during flares can be detected throughout the electromagnetic spectrum, with a significant fraction emitted in ultraviolet , especially in the form of continuum radiation \citep{Milligan}. Despite being a significant portion of the flare energy budget, the exact energetics and physical mechanisms of continuum emission is still under debate.\\

In the late 20th century, flares showing excess emission in visible wavelengths relative to the quiescent solar spectrum were termed white-light flares \citep[WLFs,][]{Najita, hudson}. Although initially thought to be rare, later statistical studies based on full-disc solar fluxes suggested that WLFs may be more common.
For example, \citet{Kretzschmar} concluded that the majority of flares show some level of white-light emission. These enhancements are not confined to the optical range; continuum brightening has also been detected in the ultraviolet (UV) and infrared (IR) regimes and, for a few cases, found to be temporally correlated with soft and hard X-ray emission observed by instruments such as GOES, RHESSI, and Fermi-GBM \citep{kleint_2016,B_flare_enh}. With improved spatial and temporal resolution, a few studies \citep{kleint_2014, Zuccarello, B_flare_enh} have demonstrated the detectability of Near-UV(NUV) continuum enhancements in specific flare occurrences using the Interface Region Imaging Spectrograph \citep[IRIS,][]{iris_instrument}. Notably, for a strong flare, \citet{kleint_2016} found that NUV enhancements tend to be stronger than those in the visible or IR and that hard X-ray (HXR) sources and the locations of NUV continuum enhancement overlap, suggesting that the emission arises at flare footpoints where energy deposition occurs. However, the general occurrence and variability of these enhancements across different flare classes remain poorly understood. In contrast to the relatively extensive statistical investigations of WLFs, no study has attempted to quantify the prevalence of NUV continuum enhancements and their dependence on flare properties. 

To derive robust statistics on NUV continuum enhancements across flare classes, the reliability of detection methods is crucial. In the visible range, white-light continuum enhancements were historically observed only in the most energetic flares (M- and X-class). Only recently have such enhancements been detected in lower-energy flares, including those of C-class intensity \citep{jess2008,WLF_linear_interpol,wlf_optimisation}.
Detection techniques range from image differencing to intensity thresholds.

\begin{figure*}[ht!]
        \centering
        \includegraphics[width=\textwidth,height=0.5\textwidth]{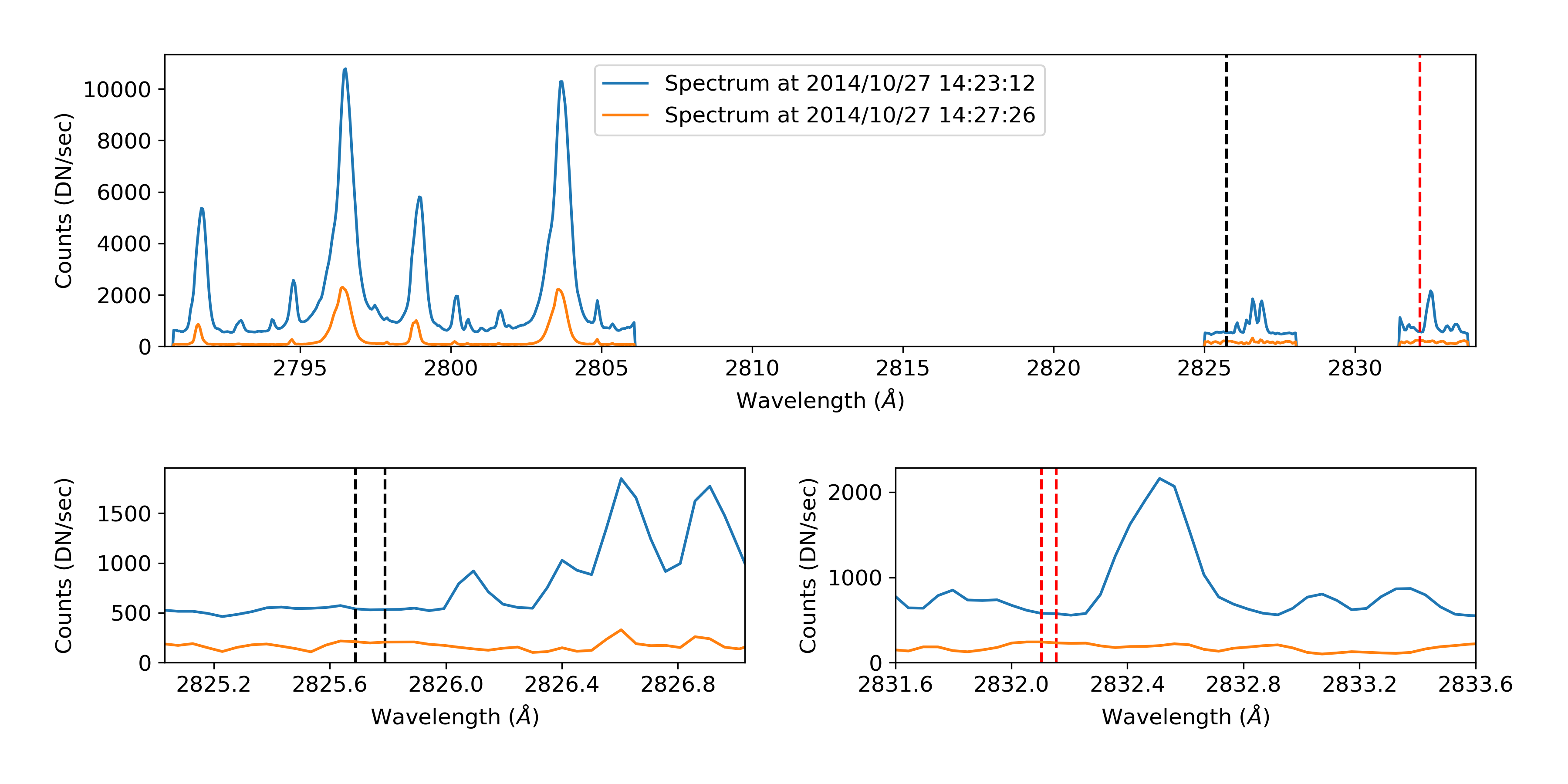}
        \vspace{-10mm}\caption{Observed Mg II h\&k spectrum, with the continuum windows used in this study 
        indicated with dashed vertical lines. The continuum windows are located well away from the wings of the strong resonant Mg II lines at 2796 and 2803 \AA.
        The lower panels show magnified views of the selected continuum regions. The blue spectrum corresponds to the time when the continuum emission was enhanced, while the orange spectrum represents the post-enhancement period, still during the flare's impulsive phase.}  
          \label{cont_window}
\end{figure*}

But these methods may fail to identify anomalous behaviors, such as enhancements occurring prior to the flare onset, because they further rely on assumed temporal or spatial correlations.  In this study we alleviate some shortcomings of the detection techniques used previously, by applying outlier detection on NUV and Far-UV (FUV) data paired with Gaussian process regression (GPR). GPR is highly adaptable to individual time series, and provides significance of the detections. This approach removes biases introduced for instance through thresholding or clustering criteria.

The physical mechanisms responsible for continuum enhancement remain debated. In the visible range, proposed explanations include photospheric back-warming \citep{Allred_backwarming, Machado_backwarming}, Alfvén wave pulses \citep{Alfven_pulses}, direct particle precipitation to the photosphere \citep{Najita}, recombination to hydrogen level n=3 producing Paschen continuum enhancement below the Paschen limit at 820 nm, and thermal bremsstrahlung, although the latter does not explain the observed EUV and soft X-ray emission \citep{Neidig}. 
Some of these mechanisms suggest the need for energy transport into the lower solar atmosphere. In the NUV, hydrogen recombination following the electron beam bombardment
is considered the dominant mechanism, suggesting that emission forms primarily in the chromosphere \citep{flarix_nuv_cont}. 
\citet{kowalski_Xflare} carried out radiative hydrodynamic simulation for a strong (X1.0) flare. Their results indicate that enhanced NUV continuum radiation originates from the layers hosting impulsively generated downflows (the chromospheric condensation (CC) layer), along with a heated stationary layer situated beneath this CC layer. This further strengthens the claim that the NUV emission ultimately originates from hydrogen recombination to the n=2 orbital level (Balmer continuum below the 364.6 nm Balmer limit) as the dominant mechanism.

The FUV continuum, by contrast, may involve contributions from UV line irradiance from the upper chromosphere and the transition region in order to photo-ionize Silicon atoms near the temperature minimum region \citep{FUV_enh}. An in-depth understanding of the physical origin of continuum enhancement for different flare energies ultimately requires flare simulations.

In this paper, we investigate the occurrence and variability of NUV (Balmer) continuum enhancements using 234  IRIS NUV and FUV flares observations of varying classes. By employing detection techniques that do not rely on coincident HXR signals or strict threshold constraints, we aim to characterize continuum enhancements in a less constrained manner. The paper is structured as follows: Section 2 describes the dataset, selection criteria, and detection methodology; Section 3 presents the statistical results, and Section 4 discusses the effects on flare energetics.

\section{Observations and Methods}

\subsection{Observations}
\subsubsection{IRIS}
IRIS is a UV imaging spectrograph that captures images and spectra of the Sun. 
This study utilizes IRIS Level-2 NUV spectral observations, which are calibrated in terms of darks, flatfield, geometric corrections, wavelength, and spatial alignment.

The Slit-Jaw Imager (SJI) onboard IRIS captures broadband images of the solar surface in the FUV (1335 \AA\ and 1400 \AA, each with a 40 \AA\ bandpass) and NUV (2796 \AA\ and 2831 \AA, each with a 4 \AA\ bandpass) with a field of view of 175\arcsec $\times$ 175\arcsec. 
The spectrograph consists of a single 0\farcs33-wide slit that observes the solar surface in the FUV (1331.7 – 1358.4~\AA\ and 1389.0 – 1407.0~\AA, with a 26 m\AA\ spectral resolution) and the NUV
(2782.7 – 2835.1~\AA, with a 53 m\AA\ spectral resolution), scanning a region in steps if the observation is not in sit-and-stare mode.

IRIS operates with ``line lists'' that define which part of the spectrum to record. The flare line list, which contains several NUV ``continuum'' windows, is most commonly used for flare observation programs.

\begin{figure}[!htb]
    \centering   \includegraphics[width=0.49\textwidth]{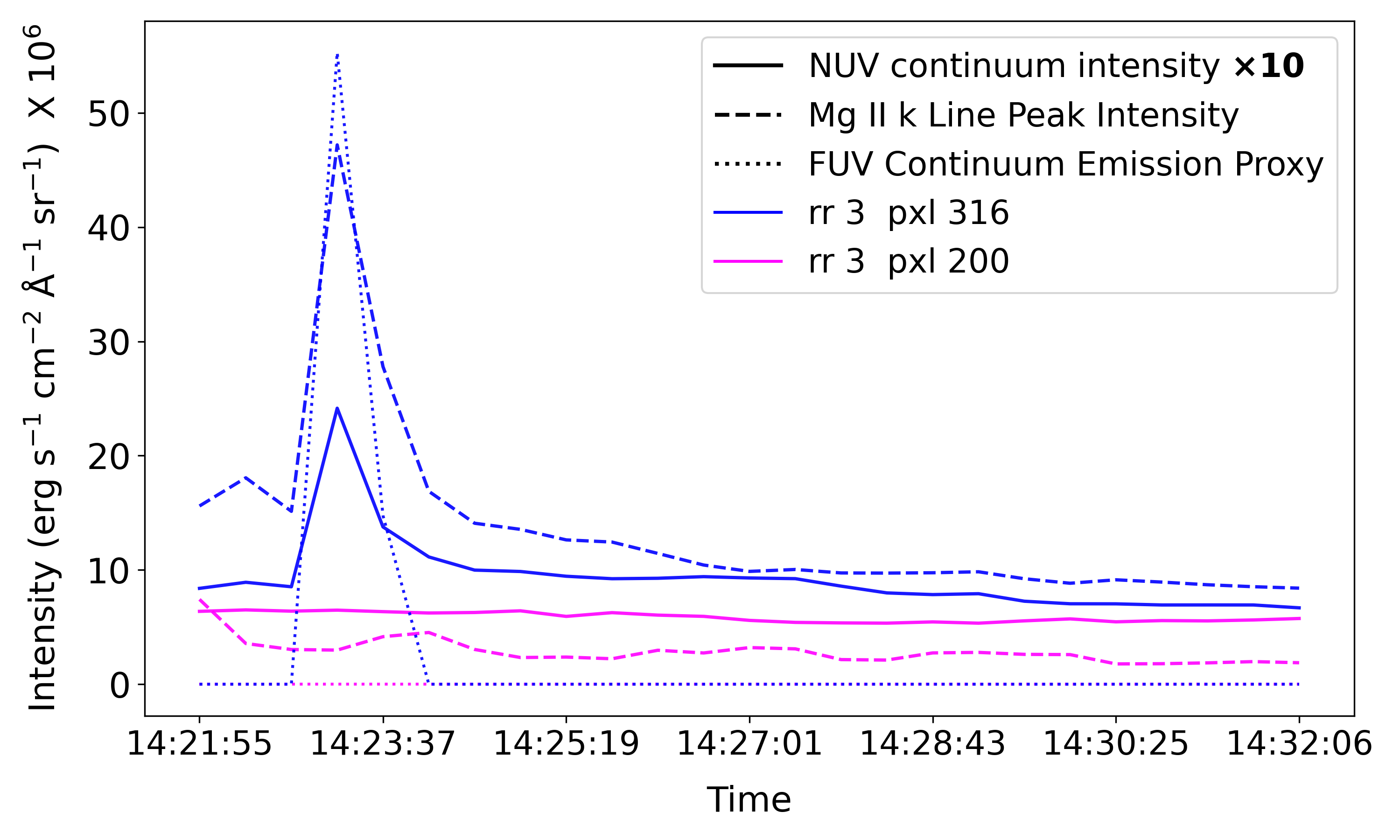}\vspace{-2mm}
    \caption{Impulsive increase observed in the NUV continuum (2826 \AA\ window; solid), the peak intensity of the Mg II k line (dashed), 
    and the FUV continuum emission proxy (dotted) at pixel (pxl) 316 (blue), raster position (rr) 3, during an X-class flare on 27-10-2014 at 14:23:12. In contrast, no such behavior is seen in the time series of other pixels on the same raster position, such as pixel 200 (pink).
    Refer to Fig.~\ref{sji_image} for the SJI image.}
    \label{enh_ts}
\end{figure}

The flare observations for this study are selected using the Google document flare list\footnote{\url{https://docs.google.com/document/d/1TAUfuErPi0QQ7aW_KoKLlEzUttGsAyAvN5eVBWHe8CM/edit?usp=sharing}} maintained by the IRIS team.
Observations with a large number of raster positions enhance the spatial sampling of the active region. However, this comes at the cost of reduced temporal resolution. To balance these factors, we restrict our analysis to observations with sixteen or fewer raster positions. 
Additionally, we exclude limb flares (heliocentric angle $\mu < 0.4$) to avoid underestimation of flare intensities. Extended emission near the limb may appear artificially fainter if a part of the emission originates in deeper layers, or suffers from projection effects.

Applying the constraints on the raster count, heliocentric angle, and the availability of  the continuum wavelengths used in this study, results in a dataset comprising 385 flare instances.
To determine the number of flares where the ribbon intersected the IRIS slit, we conducted a manual inspection of IRIS SJI images sampled at three-minute intervals throughout the GOES-reported flare duration. 
We obtained a final dataset of 234 flare instances over the period from 3 February 2014 to 10 March 2024. This data set includes 34 B-class, 150 C-class, 39 M-class, and 11 X-class flares. Lastly, we pre-process the spectral data using a pipeline described in \citet{Zbinden_2024}, 
which removes any overexposed pixels (those with overexposure of at least 3 wavelength points) and pixels with missing signals (those with negative values).

\begin{figure}[tb!]
    \centering
    \includegraphics[width=0.45\textwidth]{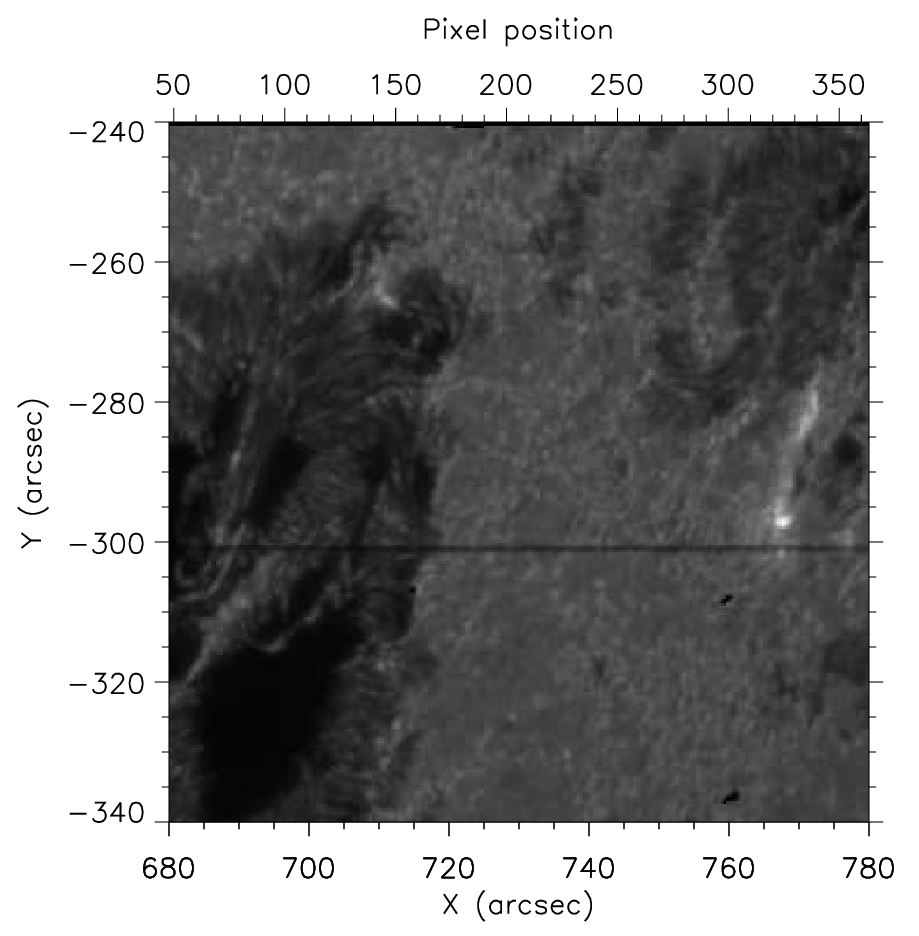}\vspace{-2mm}
    \caption{Slit Jaw Image in the 2832 \AA\ continuum window of an active region during a solar flare, showing enhanced NUV continuum emission near the center right of the image. The SJI observation time was 27-10-2014 at 14:23:16.950. 
    \label{sji_image}} 
\end{figure}

\subsubsection{GOES}
The Geostationary Operational Environmental Satellite (GOES) is a collection of geostationary satellites designed to monitor the space environment and meteorological phenomena. Since the launch of the first satellite, GOES-1, in 1975, a total of 19 satellites have been placed in orbit, with GOES 18 and 19 currently (as of 2026) in operation. 
The GOES satellites provide soft X-ray solar measurements in a long channel (1–8 \AA) and a short channel (0.5–4 \AA), with a temporal resolution of 1 s. The X-ray fluxes from GOES 8–15 were historically scaled to match the older satellites. However, it was later realized that the sensors on these satellites were already accurate and that this scaling was unnecessary. To recover the true fluxes for flares observed by these satellites, it is recommended in the GOES XRS document\footnote{\url{https://www.ngdc.noaa.gov/stp/satellite/goes/doc/GOES_XRS_readme.pdf}} that the short-channel flux be divided by a factor of 0.85 and the long-channel flux by 0.7. The flare classes available from the NOAA archival data for these older events, however, remain based on the scaled fluxes, introducing minor discrepancies in flare strength when comparing data from GOES 8–15 (pre-2020) with that from newer satellites. We therefore use the data in the following form: scaled fluxes and corresponding flare classes for events before 2020, and "true" unscaled fluxes with their corresponding flare classes for events after 2020. While this introduces some complexity when comparing pre and post 2020 flares, the scientific interpretation of our results remains unaffected, particularly because our pipeline only detected eight flares showing NUV continuum enhancements after 2020.

\subsection{Methods}
\label{sec:detection methods}  

Defining a continuum range in the 
NUV spectrum can be challenging due to the presence of spectral lines that contaminate the continuum, leading to a pseudo-continuum. 
Figure~\ref{cont_window} illustrates the continuum windows chosen for our study, the same continuum windows defined by \citet{kleint_2014}. The two narrow windows were selected to avoid contamination by spectral lines.

This continuum is not free from spectral contamination; it contains a ``forest'' of spectral lines, most of which are not strong lines. One of the windows is $2825.7-2825.8 $ \AA\ (which we will call 2826 in the following sections),
while the other is $2832.1-2832.16$ \AA\ 
(denoted 2832). 
With this definition, one can characterize an event of continuum enhancement  when there is "impulsive" emission in the NUV continuum. An example is presented in Fig.~\ref{enh_ts}. 
The figure shows a steep intensity increase, occurring simultaneously in the continuum and a spectral line. Fig.~\ref{sji_image} shows the corresponding broadband continuum emission \citep{continuumORspectral}.

\begin{figure}[ht!]
    \centering
    \includegraphics[width=0.48\textwidth]{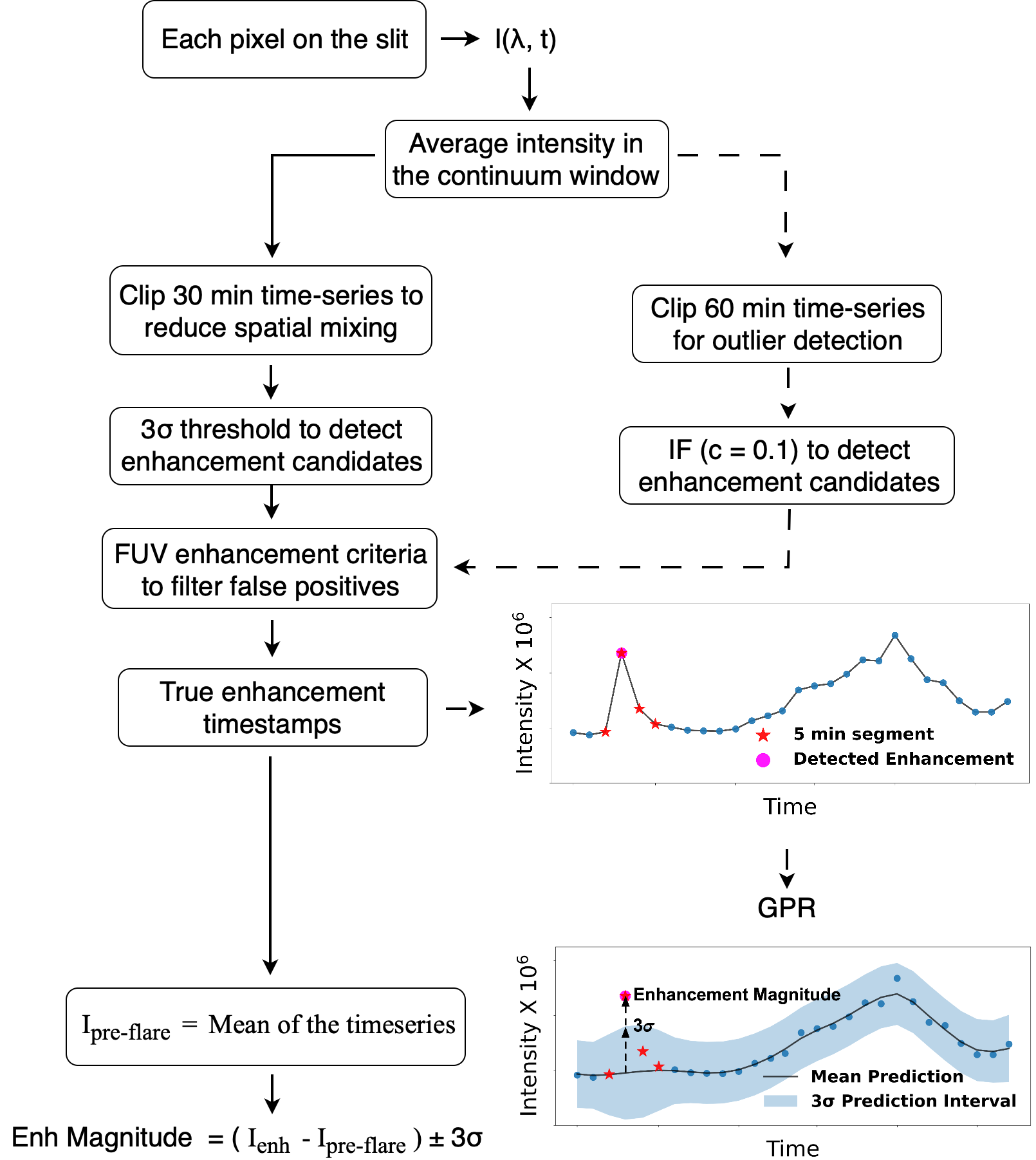} 
    \caption{Flowchart of the methodology used in this study, highlighting the differences in outlier detection and uncertainty estimation between the intensity threshold and Isolation Forest (IF) approaches. $'c'$ refers to contamination parameter which controls the proportion of detected outliers.  GPR refers to a machine learning technique known as Gaussian process regression. }
    \label{flowchart}
\end{figure}
in the center-right of the 2832 \AA\ slit jaw image. To identify such events, we construct pixel-wise time series as follows.
IRIS observations often span several hours, during which satellite drift can introduce spatial mixing, leading to artificial variations in the time series. In addition, persistent bright points tend to vary on longer timescales and may therefore contribute to gradual intensity fluctuations in the time series. Superimposed on these are shorter-timescale intensity fluctuations arising from photospheric granulation, which typically evolves over a characteristic timescale of 5–10 minutes \citep{granulation_timescale}. 
As a result, pixel-level time series show a rich diversity of variability. To mitigate the impact of spatial drift and capture granular variations on several characteristic timescales, we extracted a 30-minute segment from each time series, ideally beginning 25 minutes before and ending 5 minutes after the peak time of GOES. If the GOES peak occurs too close to the start or end of the observation, we adjust this window to maintain a total duration of approximately 30 minutes. We then tested two independent pipelines to detect continuum enhancements, estimate the pre-flare continuum level, and quantify the associated uncertainty in the enhancement magnitude. A flowchart summarizing the steps of each pipeline is shown in Figure~\ref{flowchart}. The two pipelines are described separately below.

\subsubsection{Continuum enhancement detection using intensity thresholds}

In the 2014-03-29 X-class flare, previously analyzed in detail by \citet{kleint_2016}, strong NUV \citep{kleint_2014} and FUV continuum enhancements were observed on the flare ribbons. Building on this, we examined all visually confirmed cases of NUV continuum enhancement in that event and found that they were consistently accompanied by simultaneous FUV continuum enhancements (appearing as bright bands when viewed in the pixel vs wavelength visualization for any particular timestamp) at the same spatial locations and time. We therefore tested  
the presence of this FUV signature on a subset of flares to evaluate its reliability as a necessary condition for detection in our automated pipeline. 
This filter is not intended to rule out the physical possibility of FUV-only enhancements, but rather to reduce false positives—i.e., apparent NUV enhancements that are not associated with energy deposition during a flare. This approach offers two advantages. First, it eliminates the need to assume specific spatial or temporal clustering properties of the enhancement, allowing us to detect anomalies in this behavior (e.g., identifying potential continuum enhancements occurring before the flare onset). 
Second, FUV enhancement is significantly easier to detect due to its higher contrast compared to NUV enhancement, as granulation signals are not strong in the FUV spectrum. To evaluate our method, we tested it on a subset of three flares with the following IRIS observation IDs: X-class flare at 14:04:20 20141027\_140420\_3860354980, M-class flare at 01:38:49 20141021\_181052\_3860261353, C-class flare at 17:32:00 20141019\_145935\_3860354980.
In this sample of three flares (X-, M-, and C-class), all visually confirmed instances (above 3 sigma) of NUV continuum enhancement were accompanied by co-temporal and co-spatial FUV continuum brightening, and 
all NUV-enhanced pixels with corresponding FUV enhancements were located on the flare ribbon and occurred within the GOES flare time for our flare subset. 
An example case is Fig.~\ref{enh_ts}, where one can notice the co-temporal behavior of NUV-FUV continuum enhancement. 

\begin{figure*}[!tbh]
    \centering
        \includegraphics[width=0.9\textwidth]{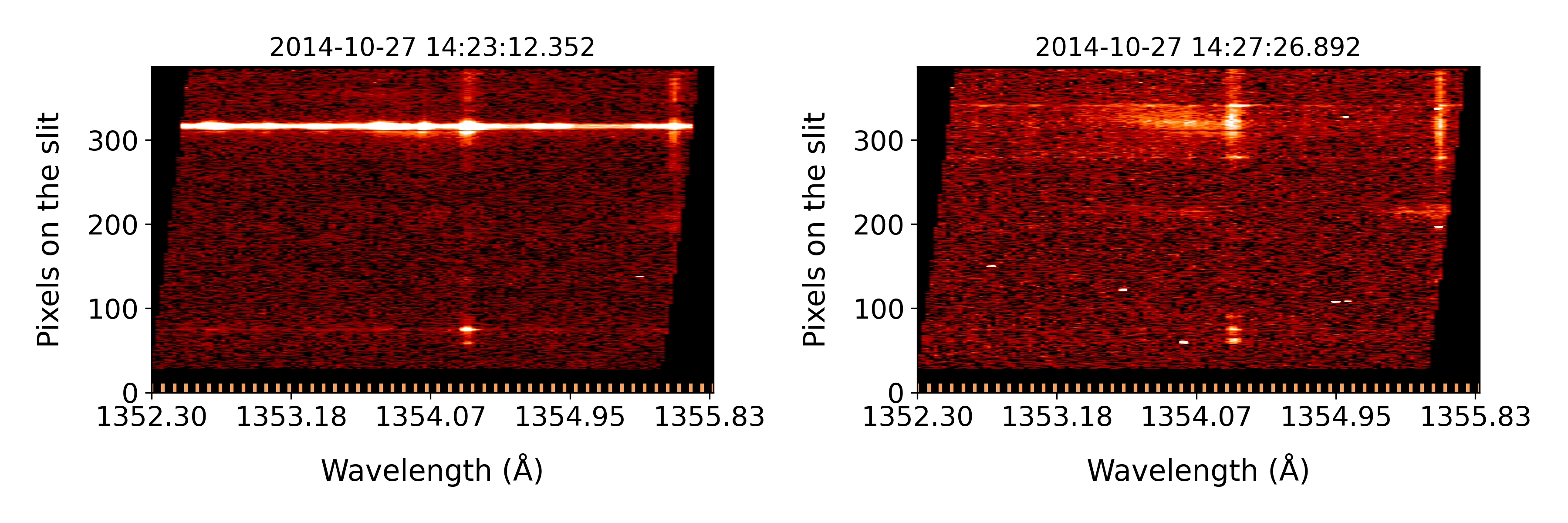}\vspace{-5mm}
        \caption{Example of continuum emission in FUV corresponding to the timestamp of enhanced continuum emission seen in NUV (refer to Fig.~\ref{enh_ts}). The right panel shows an example time-stamp where there is no continuum emission above our threshold.
        Vertical brown markers 
        along the x-axis indicate the 50 equally spaced wavelength points used in defining the FUV bright band detection criterion. }
        \label{FUV_enh_fig}
\end{figure*}
Our NUV continuum enhancement detection pipeline begins by first detecting enhancement candidates in the NUV continuum time series, followed by a filtering step that retains only those events that coincide with FUV continuum brightening. One could, of course, reverse the process and start with FUV detections. However, this would either lead to more false positives in the NUV or would make it harder to confirm the associated NUV enhancement because granulation noise is stronger in the NUV.

For the NUV continuum time series, we detect enhancements in individual time series rather than applying a global threshold. This choice is based on the observation that different pixels show varying levels of continuum variation, as also observed by \citet{wlf_optimisation}. To identify potential enhancement candidates, we detect outliers exceeding three standard deviations in each individual NUV continuum pixel time series. A pixel is detected as a potential enhancement candidate if it shows a three sigma deviation in both the NUV continuum windows. 
We subsequently developed a method to detect FUV enhancements. Since detecting relative enhancements in the FUV ($1352.30-1355.83$ $\text{\AA}$) does not require knowledge of the absolute continuum level, we do not define a fixed continuum window. 
Instead, we select 50 equally spaced wavelength points in the FUV spectrum and classify a pixel as exhibiting FUV continuum enhancement if its signal exceeds a threshold relative to other pixels at a given timestamp. Specifically, a pixel is considered enhanced if its intensity surpasses "n" times the inter-quartile range (IQR) above the third quartile at more than "m" wavelength points, effectively identifying bright bands in the FUV spectrum. The IQR-based approach requires selecting two parameters: the threshold "n" for outlier detection and the number of wavelength points "m" where the pixel must meet the threshold condition to be classified as showing continuum enhancement. As shown in Fig.~\ref{FUV_enh_fig} (right panel), spectral emission near $1354.07$ and $1355.8$ \AA\ easily satisfies $m \approx 20$. Accounting for other potential strong spectral lines, we found $m = 35$ to be neither too high nor too restrictive for detecting continuum enhancements, as there was no significant difference between detection counts for $m = 25$ and $m = 35$. However, setting $m = 45$ was too stringent for detecting bright bands. Additionally, reducing the IQR threshold did not increase the number of detections, suggesting that further FUV detections were limited by the available NUV enhancement candidates. Based on these findings, we adopted a threshold of $1.5$ IQR above the third quartile and set $m = 35$ as the criterion for detecting FUV continuum enhancement. Once a candidate passes the FUV criterion, the magnitude of NUV enhancement is calculated by subtracting the average of the NUV pre-flare continuum time-series from the enhanced value.

\subsubsection{Continuum enhancement detection using isolation forest}

 The previously described method finds events of continuum enhancements based on a 3$\sigma$ deviation in NUV time series and simultaneous emission in FUV. However, this approach may miss additional valid events since outlier candidates are marked based on the assumption that the intensity across time is normally distributed. To overcome this limitation, we introduce an algorithm based on machine learning that identifies ``deviations'' in time series and thus can expand the list of potential enhancement candidates.
 
\begin{figure}[!h]
    \centering
    \includegraphics[width=0.47\textwidth]{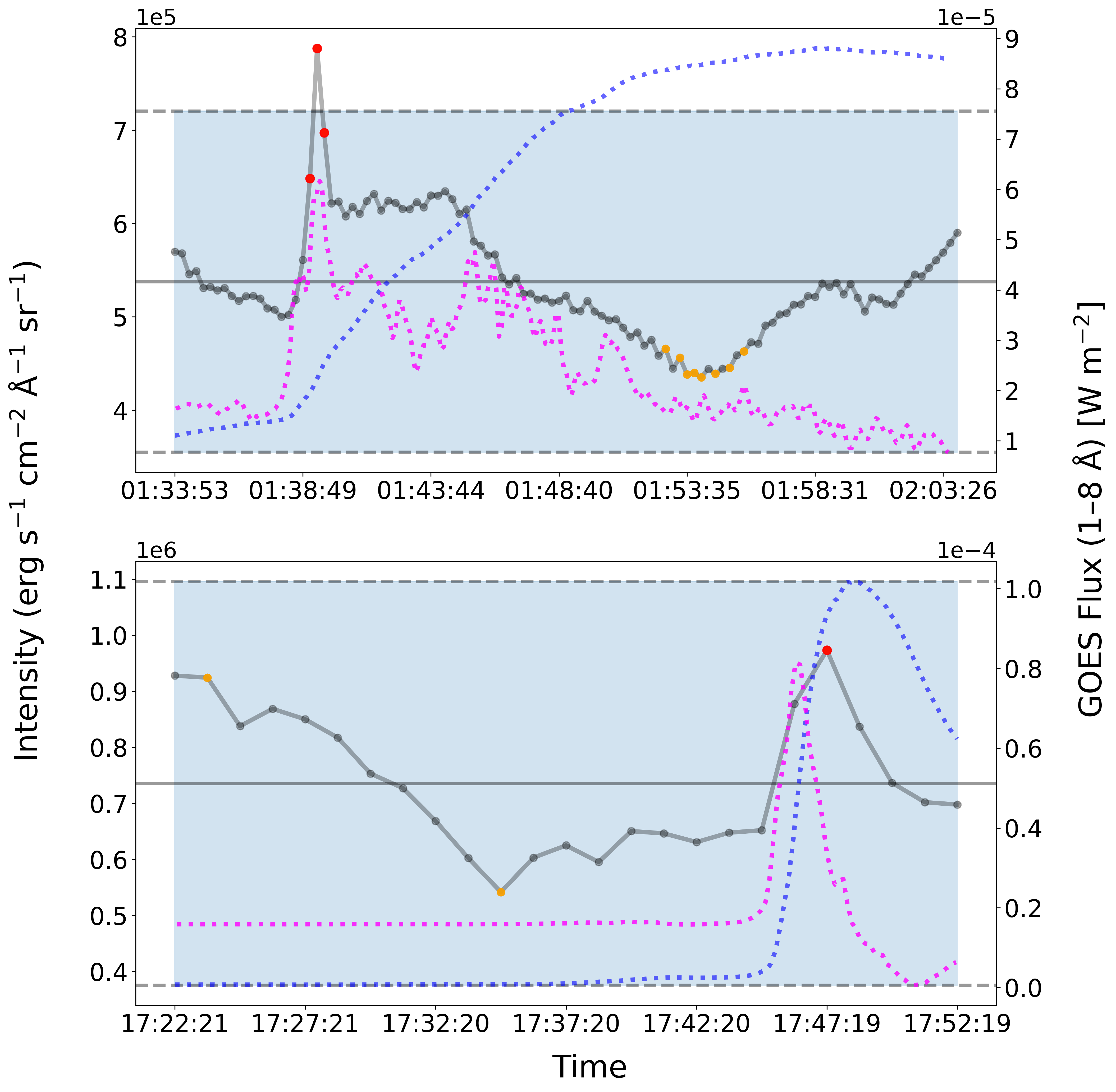} \vspace{-2mm}
    \caption{Comparison between outlier detection using three sigma threshold (blue shaded region) around the mean (black horizontal solid line) and Isolation Forest (IF) with a contamination parameter of 0.1. The colored points are outlier candidates detected by IF, and the red points are true positive enhancements filtered by detecting corresponding FUV continuum emission. The top panel (2014-10-22 M-class flare) shows that IF can better detect the rising/decay phase of the enhancement. The bottom panel (2014-03-29 X-class flare) shows the advantage of IF in detecting enhancements in cases where a simple intensity threshold would fail. The blue dotted curve is the GOES flux. The magenta dotted curve is the time derivative of the 3 sec averaged GOES flux (arbitrarily scaled) which is often used as a proxy signature of non-thermal processes.  
 }
    \label{iso_vs_nsigma}
\end{figure}

 We use the unsupervised Isolation Forest (IF) algorithm \citep{IF}, which does not assume any specific underlying distribution about the NUV continuum time-series. The algorithm works as follows:

\begin{enumerate}
    \item Randomly selects a feature (in this case, intensity values in the NUV time series, since this is a univariate time series).
    \item Splits the data along a randomly chosen threshold for that feature.
    \item Repeats this process recursively, isolating each data point from the rest of the data set.
\end{enumerate}

As illustrated in Fig.~\ref{iso_vs_nsigma}, the IF algorithm identifies not only peaks but also valleys and other distinct features. Despite detecting some irrelevant outliers (i.e., features other than peaks), we find that, with the exception of fewer than five time series, all enhancements 
found by IF that additionally pass the FUV filtering condition occur within the GOES flare event interval and on the flare ribbon. 
Moreover, we can eliminate insignificant detections when their uncertainties are quantified using Gaussian process regression, as discussed in the following paragraphs. We note that the length of the timeseries fed to the IF algorithm for outlier detection was increased to 1 hour. This was done in order to improve the uncertainty quantification performance of Gaussian process regression discussed in Section~\ref{sec:GPR CI}.
In the IF algorithm, outliers are identified on the basis of how easily the data points are isolated during the recursive partitioning process. 
\begin{figure}[ht!]
    \centering
    \includegraphics[width=0.48\textwidth]{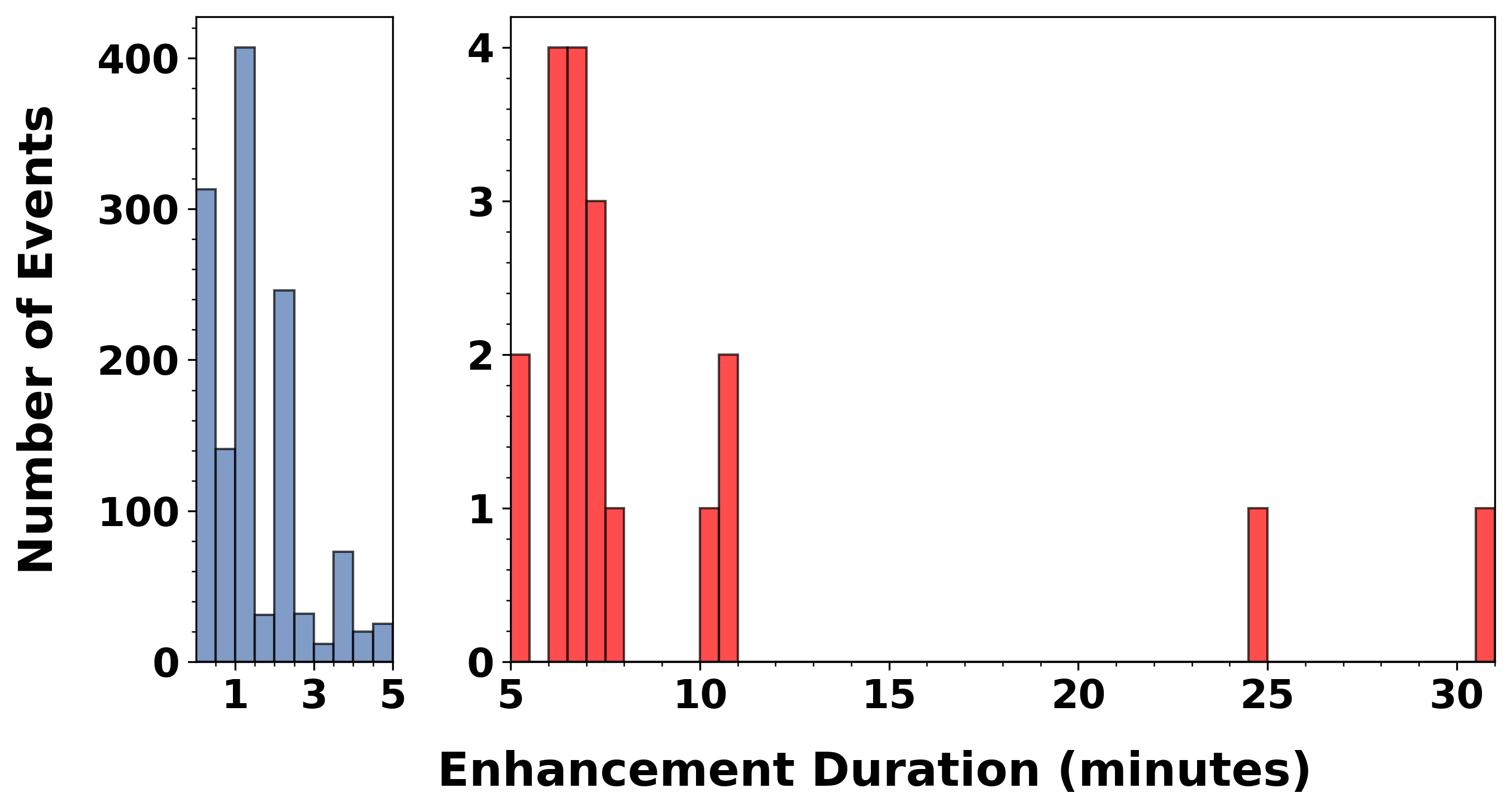}\vspace{-1mm}
    \caption{Histogram of the difference between the last and first timestamps of detected enhancement for each pixel. The majority of events last $\leq$ 5 minutes, justifying the selection of a 5-minute interpolation window for GPR. Events with durations > 5 minutes (19 cases highlighted in red) are excluded from further analysis.}
    \label{enh_dur}
\end{figure}

Data points that require fewer splits are deemed anomalous. The proportion of points classified as outliers is controlled by the contamination parameter (c), which defines the expected fraction of anomalies in the dataset. Higher contamination values result in more detected anomalies, while lower values constrain detections to the most extreme outliers. We adopt a contamination value of 0.1, corresponding to 10\% of each time series classified as outliers. 
This choice was motivated by a comparative assessment of the IF's performance against simple intensity threshold-based methods. When the contamination parameter was set to 0.05, IF yielded results comparable to those obtained using intensity thresholds. For a value higher than 0.1, IF heavily categorized inherent variations as outliers. A selection of example time series, shown in Fig.~\ref{iso_vs_nsigma}, demonstrates how the IF  more reliably captures the rise or decay phases of enhancements compared to static thresholds. Once anomalous points are detected in NUV, they are passed through the FUV filter to confirm the case of enhancement.

Since enhancements are detected on the scale of granulation variations, it is essential to estimate the underlying granulation baseline. This baseline represents the continuum intensity in the absence of enhancement and serves as the pre-flare level from which the magnitude of the observed enhancement can be derived. When the detection algorithm identifies enhancement in a pixel level timeseries, a short temporal segment (5 min, centered on the detected enhancement) is excluded from the time series to avoid contamination by flare-related emission. The remaining data is then used to construct the baseline. 

Rather than defining a single fixed baseline value, the reconstructed baseline provides a time-dependent estimate of the continuum intensity in the absence of the enhancement, such that each enhanced timestamp has a corresponding  baseline value. The enhancement magnitude is computed as the difference between the observed intensity and this baseline value. While linear interpolation could, in principle, be used to estimate the baseline \citep{WLF_linear_interpol}, it does not provide a time-dependent uncertainty estimate. In contrast, the algorithm we employ, Gaussian Process Regression \citep[GPR,][]{rasmussen2004_paper}, provides both a baseline prediction and an uncertainty estimate that adapts to the local variability of the granulation signal rather than assuming a constant uncertainty derived from the global standard deviation of the time series. We refer the reader to \citet{rasmussen2004_paper} for a detailed introduction to the algorithm and its mathematical background. We only briefly introduce the idea of the algorithm here.

\begin{figure*}[h!]
    \centering
    \includegraphics[width=\textwidth]{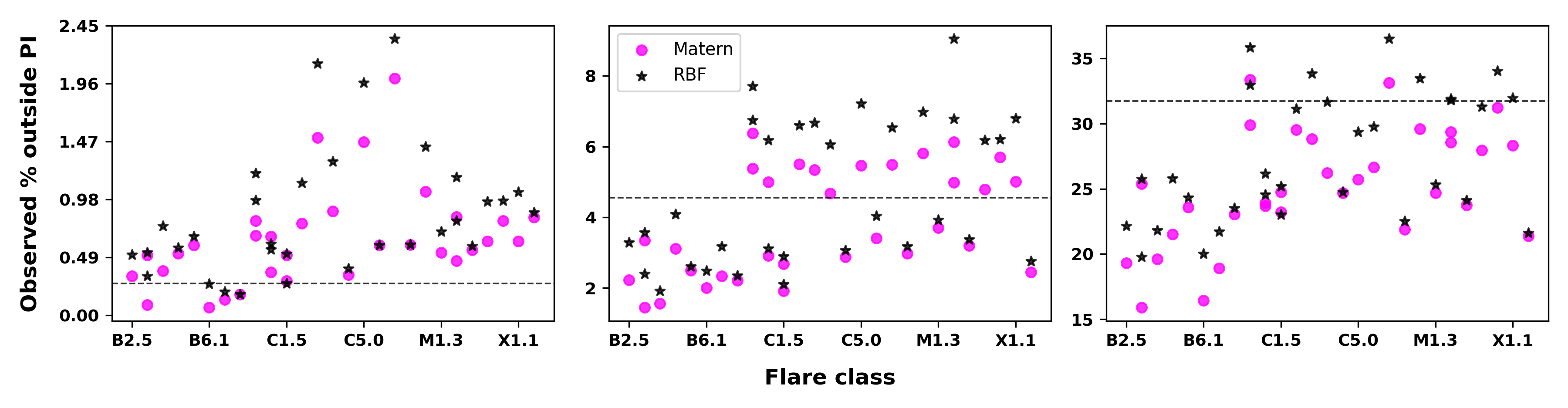}\vspace{-3mm}
    \caption{Percentage of test points outside the predicted $3\sigma$ (left), $2\sigma$ (middle) and $1\sigma$ (right) prediction interval, shown for Gaussian process models using Matérn (magenta) and RBF (black) kernels. The horizontal dotted lines represent the theoretical expected percentage of points that lie outside of the "n" sigma prediction interval. The RBF kernel overshoots the percentage of points outside the prediction interval compared to Matérn in most flare classes for the $3 \sigma$ performance evaluation, rendering Matérn kernel the better choice for uncertanity estimation. }
    \label{gpr_eval}  
\end{figure*}

GPR is a regression model popular for its ability to provide confidence intervals for each point-prediction. This regression algorithm does not simply fit an a priori known functional form to the data. Instead, it provides a probability distribution of many potential functions. These functions are characterized using a kernel, which essentially is a mathematical formula that outputs the covariance the data points would have among each other. The prior probability distribution over the functions and the errors and uncertainties are then assumed to follow a Gaussian distribution. The choice of the kernel can be guided by some prior information about the data. Once a choice is made, one can calculate the log marginal likelihood of observing the data points and maximize it to find the optimal hyperparameters associated with the kernel. Predictions can then be made using the conditional property of Gaussian distributions. To model the NUV continuum granulation time series, we use the prior knowledge that granulation signals are stochastic in nature and have a characteristic timescale. Two kernels are popular to model processes that follow a characteristic length scale: the Matern and Radial basis function (RBF) kernel.
We test both and add a white noise kernel to model any stochastic behavior in the intensity. The lower bound of the length scale was set to a value twice the cadence of the respective observation, and the higher bound was set to 5 times the cadence, as it was noted on a single test observation that smaller length scales perform better. We discuss more details of the evaluation of the algorithms' capability to provide prediction intervals in the next section.

\subsubsection{Uncertainty estimation}
\label{sec:GPR CI}  
For enhancement detection based on intensity thresholds, the uncertainty in the magnitude of the NUV continuum enhancement is estimated using the $3\sigma$ value computed from the time series itself. In contrast, for enhancements identified using the IF algorithm, the probabilistic nature of the GPR model provides prediction intervals for each point along the predicted baseline during the enhancement period. To evaluate the uncertainty quantification performance of the algorithm, we constructed an observation set comprising a mix of active regions both with and without flare events. We first excluded any pixels with continuum enhancements using the IF, followed by the FUV filtering approach, and also ensured that no major cosmic ray events were present in these observations.

This step is crucial, as spurious peaks caused by cosmic rays or continuum enhancements are unpredictable by our model and may appear in the test set, thus degrading model performance. It is important to note, however, that the presence of cosmic ray events does not affect the detection of continuum enhancement events themselves.\\

As mentioned earlier, we extended the total time-series length fed to IF to one hour to increase the number of available training points. This time-series is then fed to GPR for uncertainty quantification. This adjustment was a necessity to have reliable uncertainty estimation. For just a 30 min timeseries, low-cadence observations demonstrated poorer uncertainty quantification performance compared to high-cadence ones. Hence, extending the total time-series length to one hour provided more training points for low cadence observations. To evaluate the uncertainty quantification performance on each test observation, we randomly selected 15 pixels along the slit from each raster position (or 60 pixels in the case of a sit-and-stare observation). For each selected pixel time series, we performed five independent evaluations by randomly selecting a single segment to serve as the test set, while using the rest of the time series for training. 
In each evaluation, only one such test segment is used. This random selection and evaluation process was repeated five times per time series to ensure that interpolation performance is not biased by a particular test interval. 
The duration of the test segment was chosen based on the empirical length of enhancement events. Using the IF detection pipeline (IF + FUV condition), we found that most periods during which enhancements occur
last no more than 5 minutes $\pm 2 \times$ cadence, with only 19 cases (in red, see Fig.~\ref{enh_dur}) with longer or multiple enhancement periods.
Therefore, we defined the test chunk as a five-minute segment of the time series. In the 19 exceptional cases, the enhancements fell into two categories: (1) two separate enhancement events occurring more than five minutes apart, where one of the associated flares was not recognized by GOES; or (2) a single enhancement episode spanning both the impulsive and decay phases of the flare, resulting in a longer duration. 
The GPR algorithm works well when interpolating between points and much worse with extrapolations. Consequently, enhancement events occurring at the very beginning or end of the time series cannot be modeled (7 cases).\\

Finally, for each observation, we compute the cumulative percentage of test points that fall outside the predicted $3\sigma$, $2\sigma$, and $1\sigma$ confidence intervals. Ideally, for well-calibrated models, these percentages should match their theoretical expectations ($0.27\%, 4.55\%$, and $31.73\%$ respectively). This provides a statistical overview of the model’s uncertainty calibration across the entire observation, even though the prediction intervals themselves are computed for individual points. In principle, rigorous evaluation of point-wise prediction intervals would require knowledge of the true underlying function and the use of simulated datasets \citep{uncertainty_eval}. While we ultimately report $3\sigma$ uncertainties, evaluating multiple confidence levels provides a more complete picture of the model's probabilistic calibration. As seen in Fig.~\ref{gpr_eval}, at the $3\sigma$ level, the Matérn kernel generally performs better than the RBF kernel, with observed outlier rates remaining close to the expected $0.3\%$ threshold, albeit with mild over-confidence. At the $2\sigma$ level, the Matérn kernel again shows relatively good agreement with theoretical expectations, while the RBF kernel tends to be slightly over-confident. At the $1\sigma$ level, both models show under-confidence, with fewer outliers than expected. Importantly, even in the worst-performing flare class, the Matérn kernel still captures $\approx98.5\%$ of test points within the $3\sigma$ interval. Thus, we select the Matérn kernel over the RBF kernel, which gives the final functional form of the kernel as:
\begin{equation}
k_{(\nu = \frac{3}{2})}(r) = \left( 1 + \frac{\sqrt{3} r}{l} \right) \exp \left( -\frac{\sqrt{3} r}{l} \right) + \epsilon \hspace{0.1cm}, \text{ where }\epsilon \sim \mathcal{N}(0, \sigma_{noise}^2)
\end{equation}
where $r$ refers to the separation between any two data points and $l$  refers to the characteristic length-scale.\\
Notably, uncertainty quantification is significantly better calibrated for B-class flares; indicated by the low percentage of points falling outside the n$\sigma$ confidence intervals. A plausible explanation is that B-class flares typically involve smaller flaring regions, resulting in fewer time series being affected by dynamic, flare-related fluctuations. If poorly modeled segments correlate with heated regions, then smaller flare areas may inherently lead to better GPR fits.

\section{Results}

\subsection{Occurrence statistics of NUV continuum enhancements}

Using the detection methods described in Section~\ref{sec:detection methods}, we found the following:

\begin{enumerate}
    \item{Intensity Threshold Results:}\\
    Of 234 flare candidates 
   (some of which were not ideally observed by the spectrograph slit), the intensity threshold method followed by FUV-based filtering identified NUV continuum enhancements in 26 flares. These included 6 X-class, 6 M-class, and 14 C-class events. 
    All detected enhancements were above the significance level of $3\sigma$.  \\

    \item{Isolation Forest Results:}\\
    Applying the isolation forest method followed by FUV filtering yielded detections in 80 flares, consisting of 9 X-class, 22 M-class, 48 C-class, and 1 B-class events. 
    Among these, 49 flares had at least one pixel with enhancement above the $3\sigma$ threshold. The breakdown of the flares with $3\sigma$ detections
    is as follows: 7 X-class, 15 M-class, 26 C-class, and 1 B-class flare. For details of these observations, we refer to Table \ref{obs_table}.
    
\end{enumerate}

\begin{figure}[bh!]
    \centering
    \includegraphics[width=0.48\textwidth]{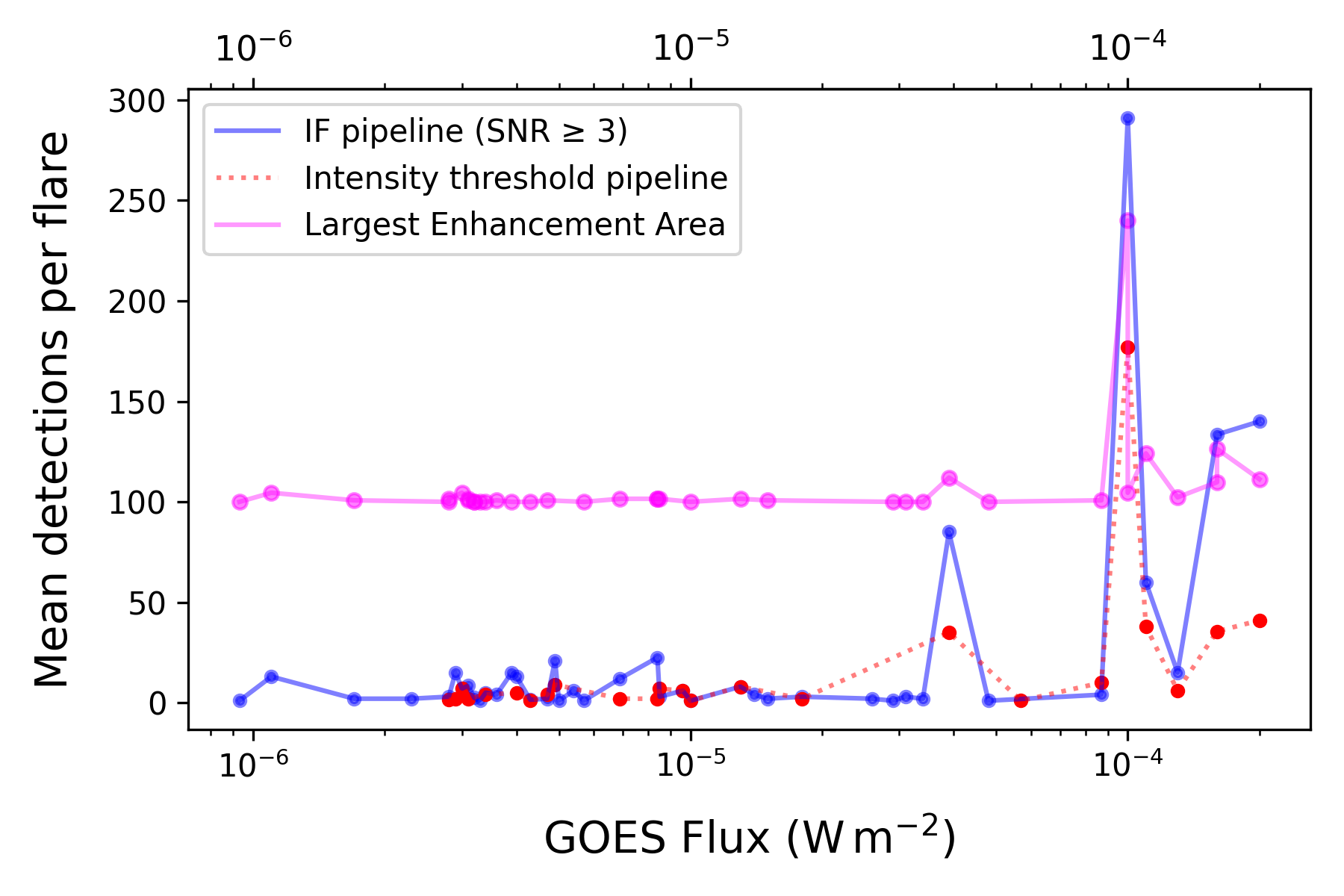}\vspace{-2mm}
    \caption{Comparison of detection performance between the IF-based ($\geq 3\sigma$) and intensity threshold-based enhancement detection methods. The y-axis shows the mean number of detected enhancement timestamps per flare, normalized by the number of flares in each GOES flux bin. The GPR-based pipeline consistently identifies more enhancement timestamps across the flare flux range. The magenta curve shows the maximum number of enhanced pixels corresponding to a full raster scan (arbitrary scaling).}
    \label{detection_comparison}
\end{figure}
\begin{figure}[tbh!]
    \centering
    \includegraphics[width=0.5\textwidth]{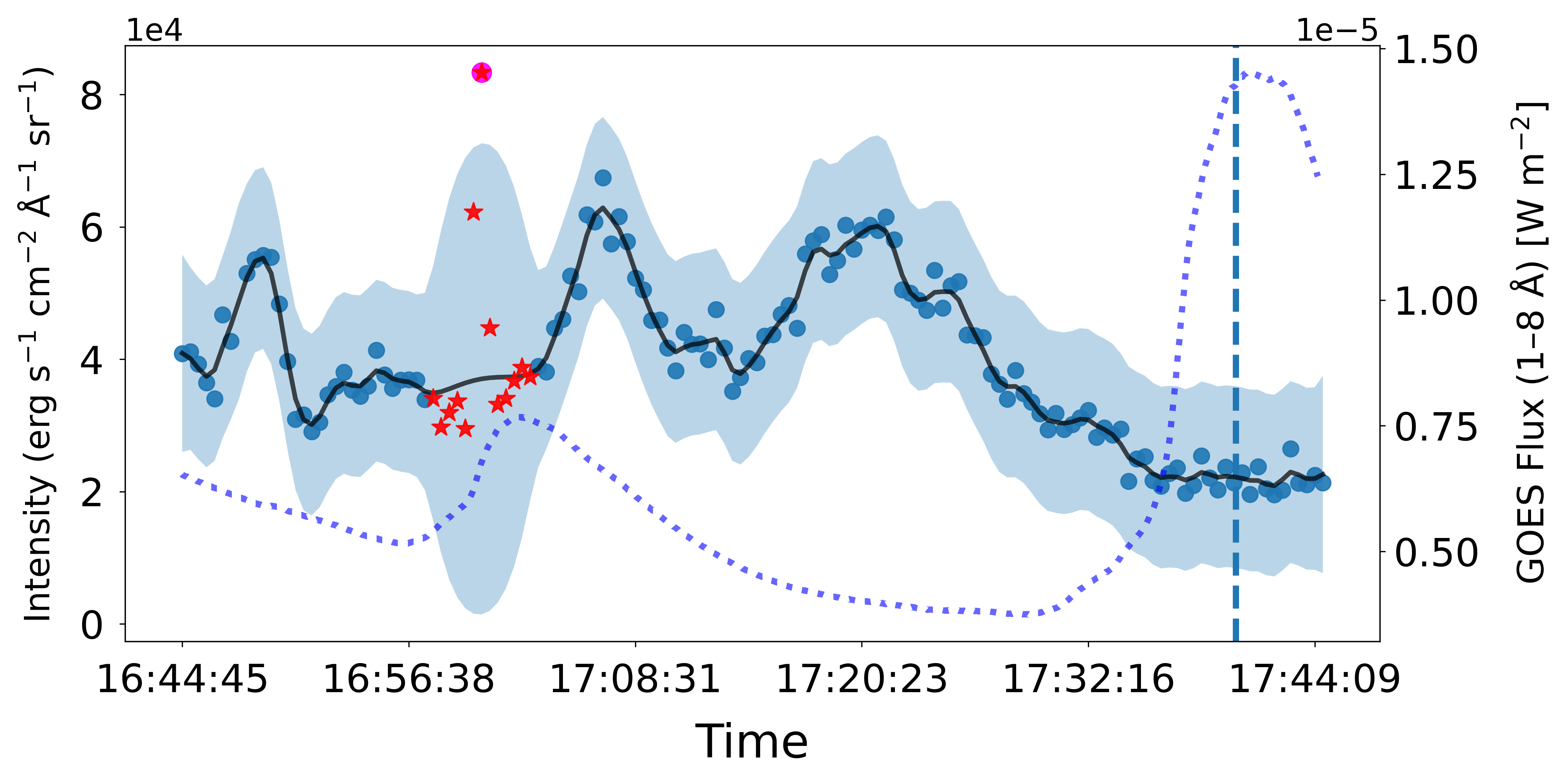}\vspace{-2mm}
    \caption{Misattributed NUV continuum enhancement caused by an unrecognized flare (27-10-2014 16:56–17:02), within a time series clipped around a GOES-reported M-class flare (blue dotted vertical line). Red stars represent the 5-minute segment excluded for GPR interpolation; the magenta circle marks the detected enhancement. The black curve and blue-shaded region represent the GPR mean and 3$\sigma$ confidence interval. GOES X-ray flux is shown as the blue dotted curve.}
    \label{double_flare_eg}
\end{figure}

\begin{figure*}[tb!]
    \centering
    \includegraphics[width=\textwidth]{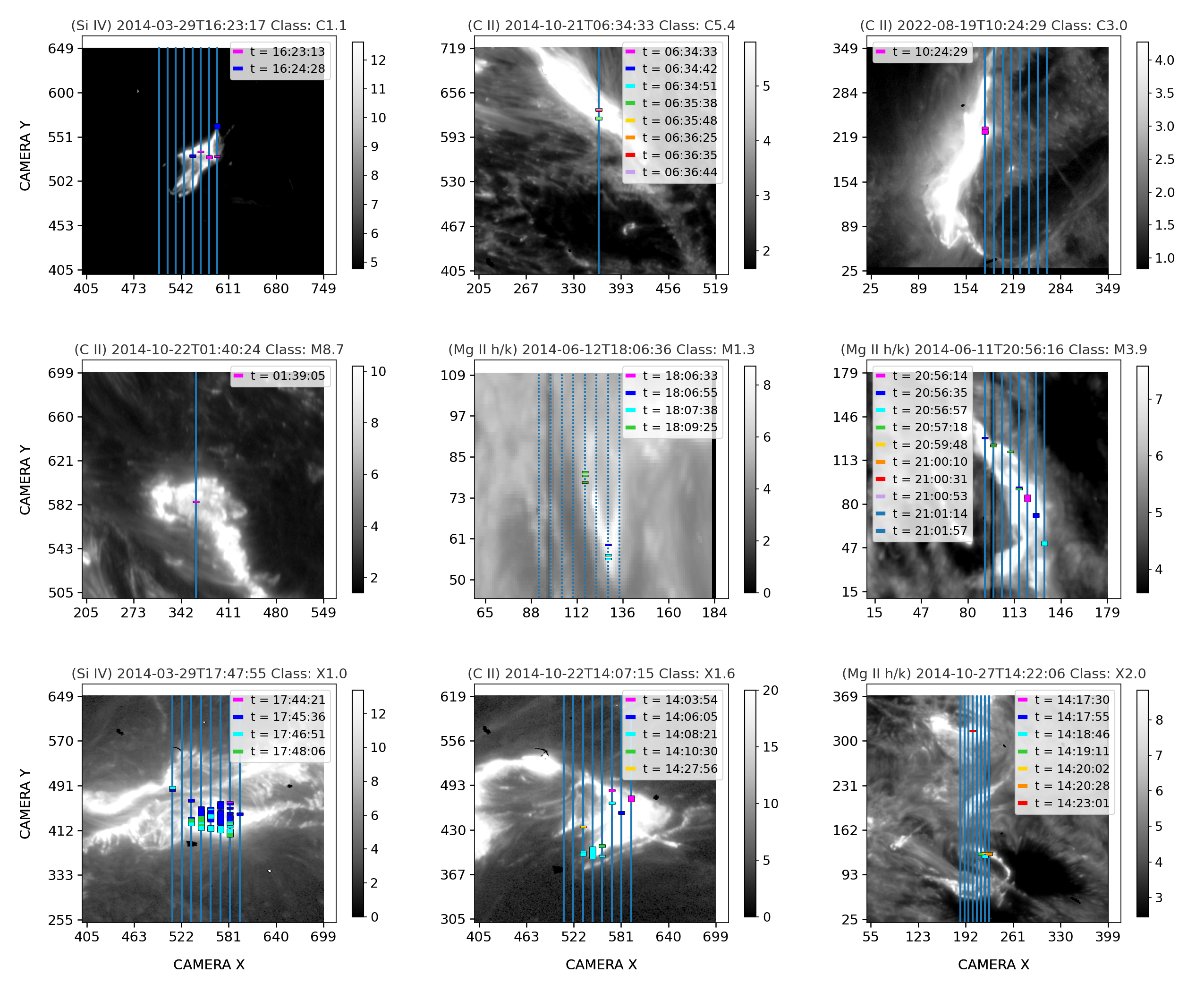}\vspace{-2mm}
    \caption{Spatial and temporal information of NUV continuum enhancements detected using the Isolation Forest pipeline. Horizontal markers show the spatial locations of significant enhancements, with thresholds set at 3$\sigma$ for C-class, 5$\sigma$ for M- and X-class, and 14$\sigma$ for the X2.0 flare to reduce visual clutter in stronger events. Thus, the area of enhancement is not directly comparable. 
    All enhanced pixels lie on the flare ribbons; any apparent spatial offsets arise from plotting multiple enhancement timestamps onto a single SJI frame. Each color represents a specific raster (i.e., all raster steps from a single raster scan), where each raster step corresponds to a different time.}
    \label{horizontal_markers_iso}
\end{figure*}

\begin{figure}[ht!]
    \centering
    \includegraphics[width=\columnwidth]{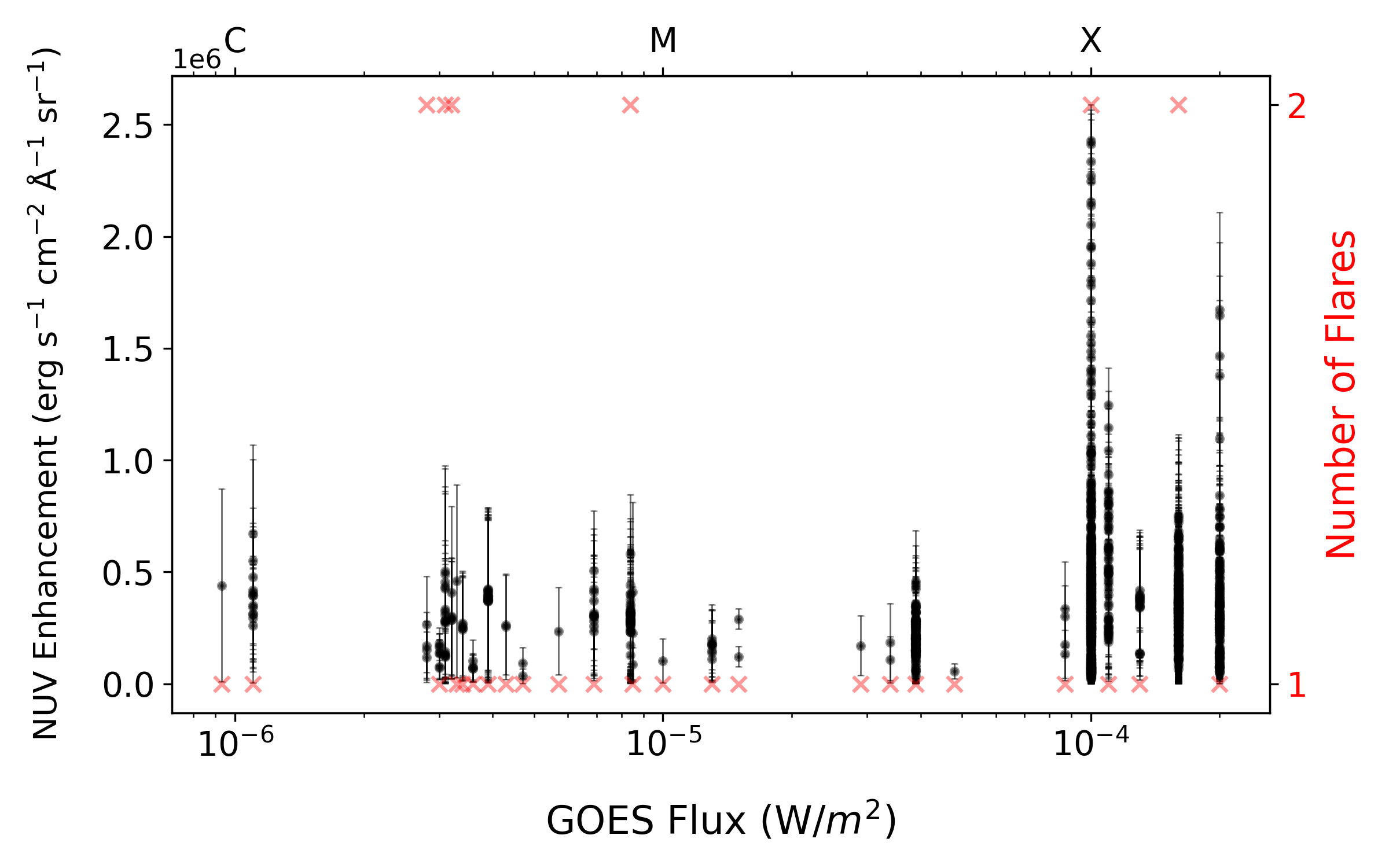}\vspace{-4mm}
    \caption{Magnitude of $\geq 3\sigma$ NUV continuum enhancements as a function of GOES flare flux, detected using the IF detection pipeline. Black points with error bars represent the enhancement magnitudes estimated from individual enhanced pixel time series for a specific flare class. A single time series may contain multiple enhanced timestamps and can therefore contribute multiple enhancement magnitude estimates to the plot. A jittered version of this plot, included in the Appendix Fig.~\ref{enh_magnitude_iso}, shows the density of detections. The total number of flares contributing to these measurements is shown by the red crosses. Stronger enhancements are predominantly observed for the most energetic (X-class) flares.}
    \label{enh_magnitude_iso_without_jitter}
\end{figure}

Based on the detection statistics, the isolation forest method detected $\geq  3\sigma$ enhancements in 20.9\% of all candidates, compared to 11.1\% using the intensity threshold method. 
The number of 3$\sigma$ detections was generally higher with the IF algorithm across flare events, and it also succeeds in detecting enhancements associated with weaker flares, see Fig.~\ref{detection_comparison} for a comparison of the detection performance of the two algorithms.
It is important to note that a few detected NUV enhancements were associated with events not listed in the GOES flare catalog (e.g., Fig.~\ref{double_flare_eg}). Consequently, a time series centered around a reported GOES flare 
may contain an unreported flare and its associated enhancement. In such cases, the enhancement could be mistakenly attributed to the GOES-reported flare, introducing ambiguity in the analysis. To ensure reliable attribution of enhancements, such ambiguous events were excluded from the statistics in Fig.~\ref{time_offset_goes}, Fig.~\ref{enh_magnitude_iso_without_jitter}, Fig.~\ref{enh_magnitude_iso} and Fig.~\ref{enh_magnitude_nsigma}. These excluded cases for the IF pipeline included  9 C-class and 5 M-class flares. 

\subsection{Spatial and temporal characteristics of NUV continuum enhancements}
 Spatial characteristics: We first report that no instances of NUV continuum enhancements accompanied by FUV enhancements were found in the quiet Sun. In the flaring regions, all enhanced pixels lie on the flare ribbons. Interestingly, these enhancements were generally observed to occur in a sub-region of the flare ribbons, as seen in Fig.~\ref{horizontal_markers_iso}. On further investigation of a few cases, the enhancements appear mostly at the edges of the flare ribbons. Figure~\ref{enh_in_subregion} shows one example demonstrating that even with relaxed significance thresholds, enhancements remain confined to the ribbon edges. 
 Therefore, there may be cases with no detected enhancements even though some part of the ribbon is observed by the spectrograph slit. The magenta curve in Fig.~\ref{detection_comparison} 
shows the largest count of enhanced pixels from among all the enhanced rasters. The arbitrary scaled values represent the largest spatial extent of enhancement during the flare, allowing for a comparison of the relative enhanced areas between different observations. It is worth noting that comparisons between the actual number of enhanced pixels cannot be made, as the spatial coverage of the flare by the IRIS slit varies significantly between observations. From our sample of 48 C-class, 22 M-class and 9 X-class flares, flares with a higher mean number of detections also exhibit larger enhancement areas at some point in time, with X-class flares showing the largest enhancement area, followed by M-class (albeit only a single flare), and then C.\\

\begin{figure*}[ht!]
    \centering
    \begin{minipage}{\textwidth}
        \centering
        \includegraphics[width=\textwidth]{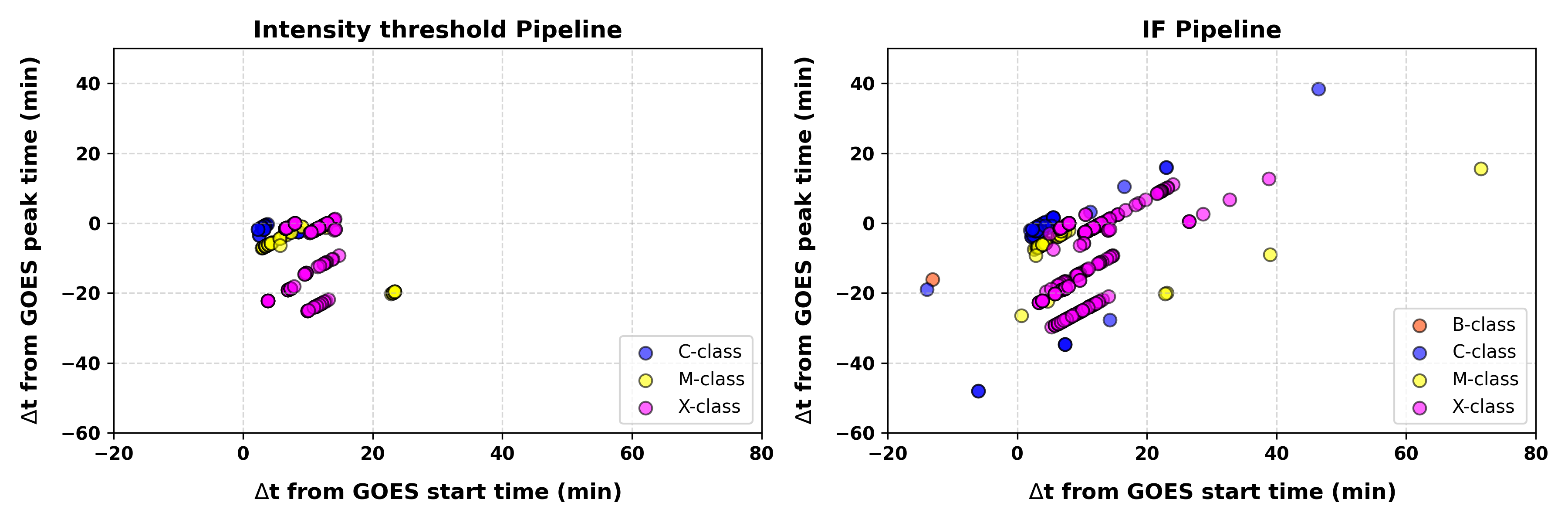}
        \caption{Time offsets of the first detected timestamp of continuum enhancement relative to the GOES soft X-ray flare start time (x-axis) and peak time (y-axis). Each point represents an individual instance (for a specific pixel) of enhancement detection, color-coded by GOES flare class. Enhancements identified by the intensity threshold pipeline (left) predominantly occur within 0–25 minutes after the GOES start time. In contrast, the IF pipeline (right) provides detections spanning a broader temporal range and detects more enhancements after the impulsive phase of the flares. This is partly because we use an hour long time series as the input to the IF pipeline in contrast to the 30-minute intensity thresholding timeseries.
        The linear patterns observed in the scatter plots arise from the detection of enhancements in temporally consecutive rasters.}
        \label{time_offset_goes}
    \end{minipage}%
\end{figure*}
\begin{figure*}[hb!]
    \centering
         \includegraphics[width=\textwidth]{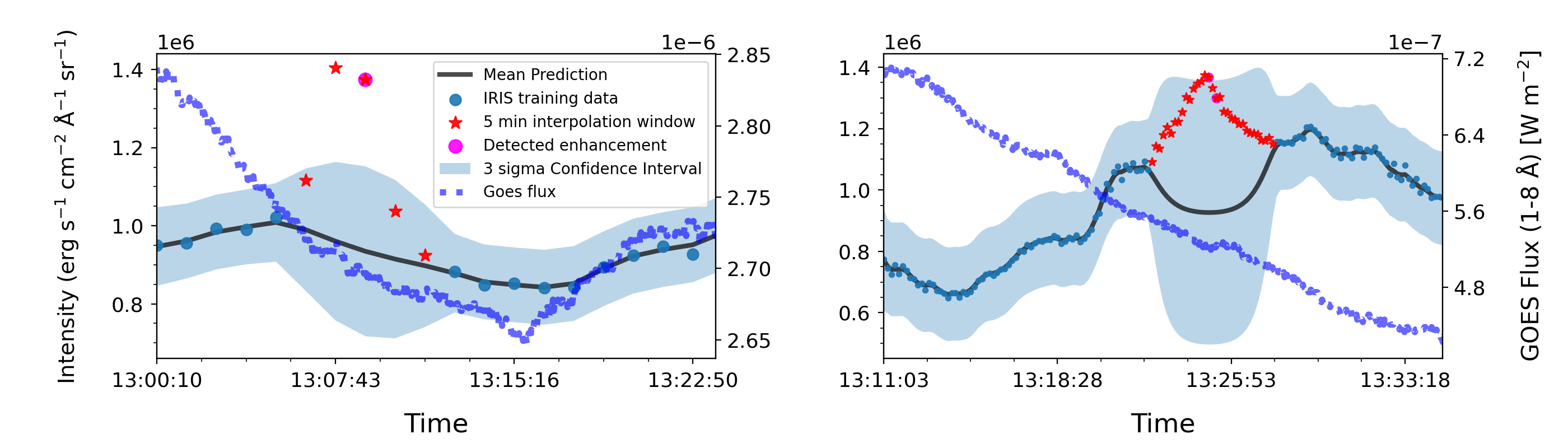}\vspace{-2mm}
        \caption{Example time-series wherein the IF pipeline detected $\geq 3\sigma$ NUV continuum enhancement with corresponding FUV continuum enhancement but without any apparent change in GOES soft X-ray flux at the timestamps of enhancement. The GOES start time for the left observation (2014-02-11 C-Class flare, raster position 5, pixel 812)  was 13:15:00, whereas for the right observation (2015-09-16 B-Class flare, raster position 0, pixel 98) it was 13:38:00. NUV continuum enhancement was detected approximately 7 min and 13 min before the GOES start time.} 
        \label{enh_without_goes_change}
\end{figure*}

Temporal characteristics: Most of the enhancements detected by the intensity threshold pipeline showed a sharp impulsive rise followed by an exponential-like decay (see Fig.~\ref{enh_ts}). 
These events were consistently associated with well-defined GOES soft X-ray flares and predominantly occurred up to approximately 25 minutes after the GOES start time and 30 minutes before the GOES peak time (see Fig.~\ref{time_offset_goes}). In contrast, the IF-based detection pipeline identified enhancements across a broader temporal range (partially because we feed it a longer time-series), ranging from 20 minutes before to 70 minutes after the GOES start time, and from 50 minutes before to 40 minutes after the GOES peak time, with most of the enhancements lasting $\leq 5$ min $\pm 2\times$ cadence (see Fig.~\ref{enh_dur}). 
In a few cases (4 pixels overall), for example, those with enhancements occurring approximately 10 and 20 minutes before the GOES start time or $>$ 20 min after the GOES peak time, enhancements were detected without any clear corresponding change in the GOES soft X-ray flux
(see Fig.~\ref{enh_without_goes_change}), unlike what is observed in other detected cases (Fig.~\ref{iso_vs_nsigma}). A visual inspection of such cases showed small-scale brightenings of a maximum of a few pixels in the SJI.

\subsection{Variation of NUV enhancements across flare class}
Figure~\ref{enh_magnitude_iso_without_jitter} shows the magnitude of $\geq3\sigma$ NUV continuum enhancements with respect to the corresponding GOES flare flux, identified using the IF detection pipeline. Black points with error bars correspond to measurements at distinct timestamps within enhanced pixel time series, wherein a single pixel may contribute multiple detections. 
The distribution clearly shows that X-class flares show both the highest enhancement magnitudes and the largest number of detection instances. Notably, 3 out of 4 X-class flares contribute more detections than all M- and C-class flares combined (see Fig.~\ref{enh_magnitude_iso}). 

While some C-class flares show enhancement magnitudes that approach or exceed those in the M-class, the associated uncertainties ranging up to 100\% in some cases, limit the ability to draw statistically significant conclusions. The C-class enhancements show larger relative uncertainties, while M-class flares display enhancement with tighter error bars. A trend of increasing enhancement magnitude with flare class is evident in the detections made using the intensity threshold pipeline (see Fig.~\ref{enh_magnitude_nsigma}). Some flares show notably fewer and weaker detections, likely because the IRIS slit did not sufficiently cover the flare ribbons. Enhancement detections in the lower flare classes are more uncertain, likely due to intrinsic flare weakness and low signal-to-noise ratio. Only a single pixel corresponding to a B-class event satisfied the detection criteria and was therefore excluded from a statistical interpretation. The concentration of detections at higher flux levels may be attributed to the larger enhancement areas of stronger flares. (see Fig.~\ref{detection_comparison}).

\section{Discussion}

\subsection{Location of NUV enhacements on flare ribbons}
Spatial analysis of continuum enhanced pixels in events such as those shown in Fig.~\ref{horizontal_markers_iso} reveals that continuum enhancements are confined to localised regions of the flare ribbon, specifically the ribbon edges, see Fig.~\ref{enh_in_subregion}. In the WLF detection regime, enhancements are reported to be concentrated near the central regions of flare kernels, which were attributed to preferred energy deposition in the central regions \citep{enh_in_subregion}. This difference in the location of NUV enhancements and visible continuum enhancements might be attributed to the lower spatial resolution of HMI or to the use of pseudo-continuum intensity estimates \citep{hmi_pseudocont}. However, the location need not be the same if the physical mechanism of the enhancement differ for the two wavelength regimes.
Because flare ribbon fronts mark the chromospheric footpoints of newly reconnected field lines and our detections lie near ribbon edges, this raises the possibility that the enhancements are co-spatial with the ribbon-front locations exhibiting the peculiar \ion{Mg}{ii} line profiles \citep{ribbon_front_brandon}, \ion{He}{i} 10830 \AA\ dimming \citep{he_dimming}, and \ion{Ca}{ii} and \ion{He}{i} profiles showing transitions between up-flows and down-flows on timescales of order 10 s \citep{ribbonFront_UpflowsDownflows}. 
In terms of flare energetics, this would mean that the enhancements occur predominantly in the regions with freshly reconnected magnetic field lines. Recent simulations of \ion{Mg}{ii} emission at the ribbon front suggest that the high-energy flux injected during the ribbon-front lifetime is relatively modest and more gradual compared to the flux injection in the regular bright ribbon regions. These ribbon-front areas exhibit weaker chromospheric evaporation than the bright ribbon regions \citep{Ribbon_front_graham}. 
Indeed, a visual inspection of a few spectra from the locations with enhancements show peculiar \ion{Mg}{ii} spectra similar to the ribbon front categories found by \citet{ribbon_front_brandon}, but we also found other types of flare spectra whose analysis will be conducted in future work. Such a study would help understand if regions with relatively modest flux of high-energy particles and gradual injection (simulated as a triangular injection profile of electron beams reaching its peak flux at 10 sec, instead of a constant flux profile, see \citep{Ribbon_front_graham}) is the preferred site for NUV continuum enhancements.

\subsection{Comparison of NUV and WL continuum emission}
Our results demonstrate clear differences in the performance of the two detection pipelines. 
The IF pipeline achieved more detections than the intensity-threshold method, suggesting that a significant number of genuine NUV continuum enhancements are missed by thresholding applied to IRIS data. However, the overall detection rate remains lower than that reported for white-light flare (WLF) studies \citep{WLF_linear_interpol, wlf_optimisation}, wherein the detections reach a rate of $> 50\%$ of the events.
One reason for the discrepancy may be that visible detections were done using pseudo-continuum intensity estimates \citep{hmi_pseudocont} obtained with an imaging instrument (HMI), while our NUV detections rely on a slit spectrograph, where it is more difficult to capture the time and location of the enhancement. This reasoning is supported by our results showing that these enhancements are predominantly concentrated in the flare ribbon edges, hence it is important not only to capture the flare ribbon but also to observe the specific subregion of the ribbon where the continuum enhancement occurs.

In terms of temporal evolution, we observe that NUV continuum enhancements occur primarily during the impulsive phase of the flares but for some strong flares, also occur up to 20 min after the GOES peak time. Traditionally, enhancements aligned temporally with the impulsive phase have been interpreted as signatures of non-thermal electron beam heating and associated hydrogen recombination in the chromosphere.  \citet{enh_before_flare} however, reported a M4.5 and a M1.6 WLF whose white-light emission peak occurred $\approx$10 min after the soft X-ray maximum. These enhancements were hypothesised to be associated with the cooling phase of the flare. Our results also suggest that enhancements are not limited to just the impulsive phase. We speculate about the possibility of multiple episodes of energy deposition, potentially involving repeated electron beam bombardment \citep{multiple_plasmoids}.

Moreover, a few single-pixel NUV continuum enhancements (3-4 cases) did not correlate with any clear change in the soft X-ray profile, leading us to speculate that they correspond to the decay-phase from earlier flares or small reconnection events in the solar atmosphere.
In reference to Fig.~\ref{enh_without_goes_change}, there is no GOES reported flare preceding the C-class flare. For the B-class flare, a C-class event occurred immediately beforehand. These detections corresponded to small, spatially localized bright points that were structurally indistinct from other features in their vicinity. These cases look exactly like flare-related NUV enhancements in the NUV and FUV continuum, but appear very rarely (4 cases in the 250 analyzed flares) and are limited to single pixel detections on the IRIS slit.

\subsection{Spatial correspondence between NUV and FUV enhancements}

We find spatial and temporal correlation between NUV and FUV continuum enhancements during flares. We discovered that strong NUV continuum enhancements were accompanied by FUV continuum enhancements, however, this does not rule out the possibility of the two occurring independently.  
Radiative Hydrodynamic simulations of strong flares have shown that the recombination of free electrons with protons, following hydrogen ionization by non-thermal electron beams, can account for the NUV continuum emission \citep{flarix_heinzel,kowalski_Xflare}. In contrast, the FUV continuum near 1520 \AA\ and 1680 \AA\ is believed to arise from the recombination continua of neutral silicon. These continua are believed to be produced when neutral silicon atoms near the temperature-minimum region of the lower chromosphere, are photo-ionized by intense flare UV line emission originating from the upper chromosphere and transition region (e.g., \ion{C}{ii}, \ion{C}{iv}). The subsequent recombination of electrons with \ion{Si}{ii} ions emits the observed FUV continuum (Doyle \& Phillips 1992). Flare simulations have not yet compared the formation mechanisms of the FUV and NUV continua. We plan to investigate this in future work using hydrodynamic simulations of the flare candidates in our study.

\section{Conclusions}

In this study, we developed and applied two continuum enhancement detection pipelines to IRIS spectral data. For one pipeline, we used a machine learning algorithm to obtain the uncertainties over enhancements, which allowed us to detect enhancements within the granulation levels in flares as low as C1.1 class. Below we list our main findings :
\begin{enumerate}
    \item Based on the analysis of 243 flare cases, 3$\sigma$ significant NUV continuum enhancements were detected using the Isolation Forest detection pipeline in 49 flares, which is twice the number of detections obtained using the intensity threshold approach. This number appears relatively low compared to continuum enhancements detected in imaging data in the WLF regime, primarily because IRIS is a slit spectrograph.
    \item NUV continuum enhancements occur predominantly on the flare ribbon edges. This suggests the possibility that the enhancements occur in regions with freshly reconnected magnetic field lines, or the ribbon fronts. There have been previous findings of peculiar \ion{Mg}{ii} line profiles, short-lived upflows, and \ion{He}{i} dimming at the ribbon fronts during flares. Whether NUV continuum enhancement preferably occurs at the location of the above-mentioned features is to be verified in future work.
    \item The detected enhancements are stronger for more energetic flares and occur predominantly in the impulsive phase of the flare but also up to 20 min after the GOES peak time for strong flares. This suggests the possibility of multiple reconnection events. There were also 4 cases of single pixel enhancement detections that did not correspond to any clear change in GOES soft X-ray flux. These enhancements were located on morphologically indistinct bright points in the active regions and could signify small-scale energy deposition events.
    \item Finally, NUV and FUV enhancements are observed to correspond spatially and temporally. This finding allowed us to distinguish NUV continuum enhancements from other variations in the continuum time series. Future studies planning to simulate either of the continua could use the spatial and temporal correspondence as a constraint to the model. Understanding the physical origin of this correspondence is planned for future work.
\end{enumerate} 
The strength of enhancements derived in this study will help us constrain the flare simulation models not just for strong flare events, as done in previous studies, but for different flare energies. The beam parameters in the flare simulations govern the penetration depth of the electrons into the solar atmosphere \citep{RADYN}.
A clear progression from this work is to understand the relation between the flare beam parameters obtained from hard X-ray data, and the observables obtained for NUV continuum enhancement from this study.
While previous studies focused more on strong X-class events, we aim to understand whether the same underlying process for strong flares can also explain the observed enhancement in weak flares.

\begin{acknowledgements}
     This work was supported by a SNSF PRIMA grant and a SERI-funded ERC CoG grant. We are grateful to LMSAL for allowing us to download the IRIS database. IRIS is a NASA small explorer mission developed and operated by LMSAL with mission operations executed at NASA Ames Research Center and major contributions to downlink communications funded by ESA and the Norwegian Space Centre. This research has used NASA’s Astrophysics Data System Bibliographic Services. Some of our calculations were performed on UBELIX (http://www. id.unibe.ch/hpc), the HPC cluster at the University of Bern. We thank the referee for his/her valuable comments and suggestions. We are grateful to our colleagues from the space weather group at the University of Bern: Dr. Michelle Galloway, Vanessa Mercea, Moritz Meyer zu Westram and Dorian Paillon for helpful discussions that contributed in shaping the methodology of this study. The authors also thank Dr. Phil Judge for helpful discussions on the interpretation of this work.

\end{acknowledgements}

\bibliographystyle{aa}
\bibliography{references.bib}  

\begin{appendix}
\onecolumn
\section{Additional figures}

\subsection{Enhancement at the ribbon edges}
\begin{figure}[h!]
    \centering
    \includegraphics[width=\textwidth]{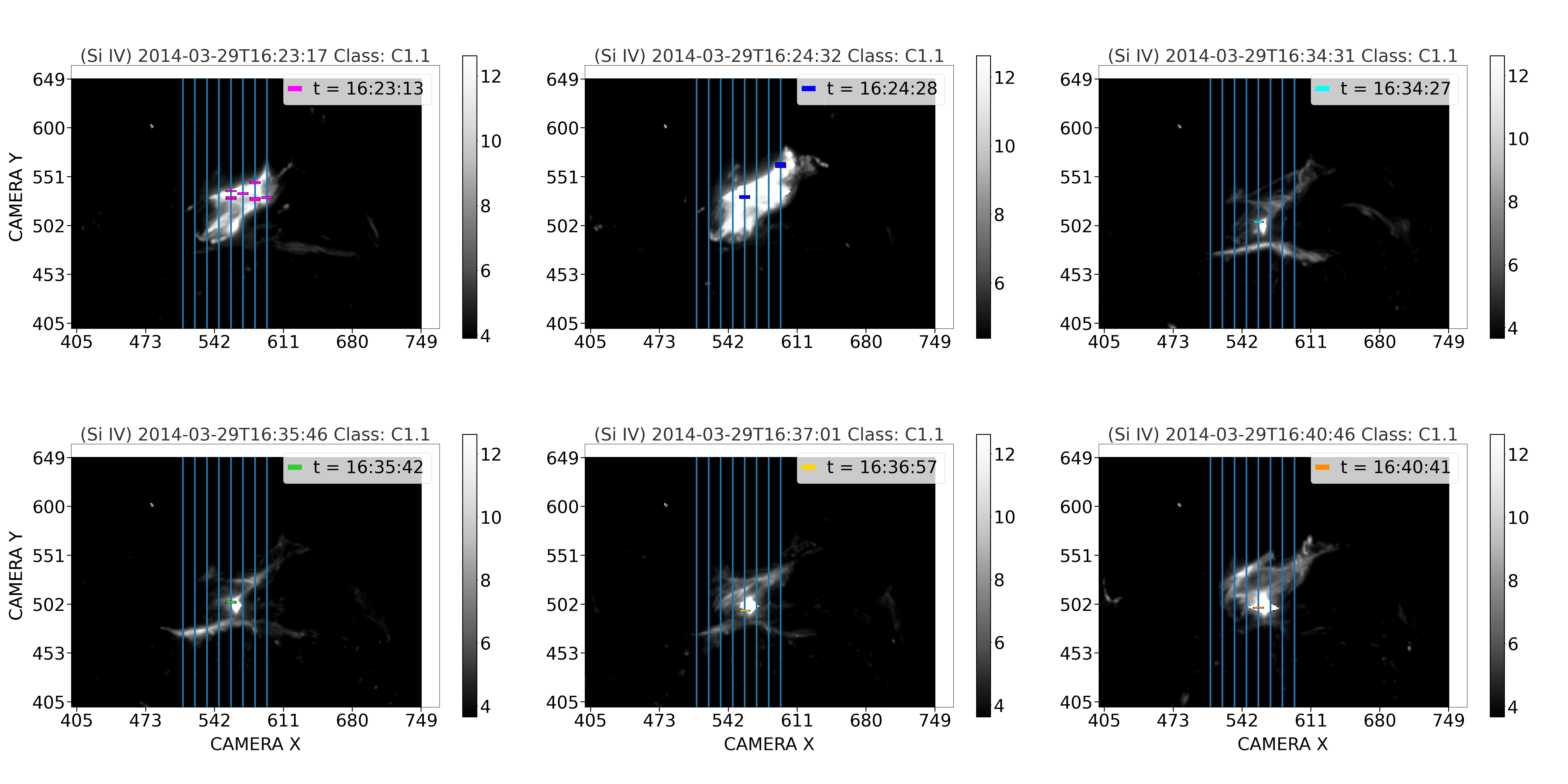}
    \caption{Spatial and temporal characteristics of NUV continuum enhancements detected using the Isolation Forest pipeline. Horizontal markers show the spatial locations of NUV enhancements, without applying any significance threshold. Although there is a time difference between the SJI image timestamp and the earliest NUV enhancement detection (plotted in magenta), the flare ribbon remains relatively unchanged in size between 16:22:58 and 16:23:17. Therefore, the magenta enhancement lies on the edge of the flare ribbon. After 16:25:28 the ribbon begins to shrink in size.}
    \label{enh_in_subregion}
\end{figure}

To avoid clutter while still conveying the spatial distribution of NUV continuum enhancements, stringent sigma thresholds were applied in Fig.~\ref{horizontal_markers_iso}. However, to more clearly illustrate that these enhancements occur on the edges of the flare ribbon, it is necessary to show all detected enhancements regardless of their significance and not have them overlaid on a single SJI frame. Figure~\ref{enh_in_subregion} provides such an example. For the SJI timestamp 16:23:17, the first instances of continuum enhancement are marked in magenta and occurred between 16:23:13 and 16:24:28. Although there is a cadence mismatch between the SJI images and the raster steps—raising the possibility that the ribbon was larger before 16:23:17 and the enhancements initially spanned the entire ribbon. The progression of the flare ribbon was examined to address this ambiguity. We find that the ribbon continued to grow in size until 16:25:28, and that it had approximately the same spatial extent between 16:22:58 and 16:23:17. Therefore, we conclude that the detected enhancements at that time were confined to the edge of the flare ribbon.

\subsection{Magnitude of enhancement: detections from both pipelines}
\begin{figure}[h!]
    \centering
    \includegraphics[width=0.83\textwidth]{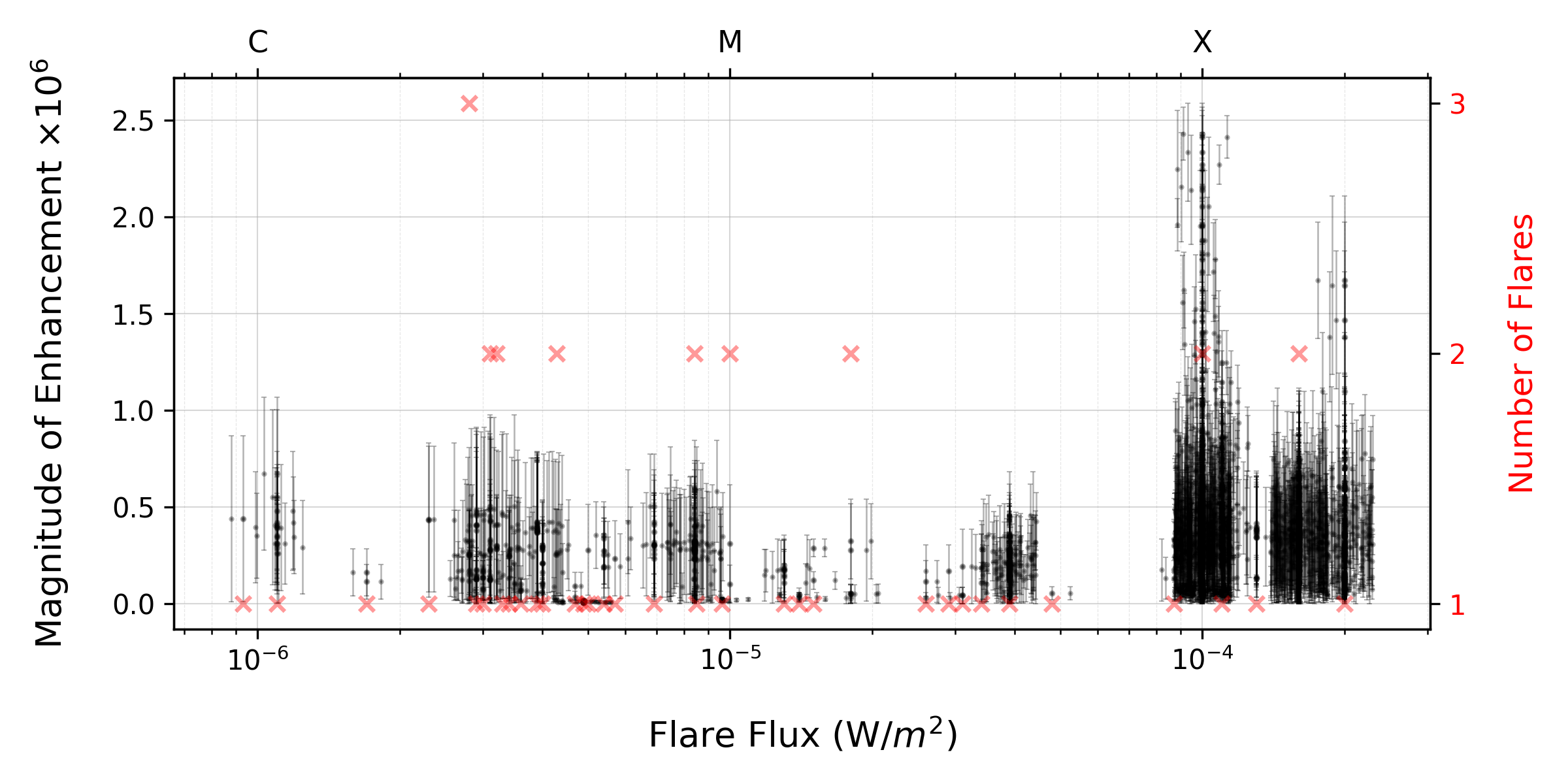}\vspace{-3mm}
    \caption{Magnitude of $\geq 3\sigma$ NUV continuum enhancements across different GOES flare flux, detected using the Isolation Forest (IF) pipeline. Small random offsets (jitter) have been applied to the flare flux axis to reduce overlap and improve visualization. The marker definitions are consistent with those used in Fig.~\ref{enh_magnitude_iso_without_jitter}. The introduction of jitter helps show a clear trend of increasing enhancement magnitude within the M-class flare regime. However, a few detections from X1.0-class flares may overlap with the M9.9-class region due to the jitter, which is why Figure~\ref{enh_magnitude_iso_without_jitter} should be consulted in parallel for accurate interpretation.}  
    \label{enh_magnitude_iso}
\end{figure}

Figure~\ref{enh_magnitude_iso_without_jitter} presents the variation in the magnitude of NUV continuum enhancements across flare classes but it lacks the resolution to reveal trends within individual classes (e.g., across all M-class flares) and does not convey the density of detections. To address this, Figure~\ref{enh_magnitude_iso} includes small random offsets (jitter) applied to the flare flux values to reduce over-plotting and improve visualization. The jitter is introduced by multiplying the original flare flux by $10^{0.06}$. This perturbation is small enough to preserve the broad separation between C-, M-, and X-class flares, though slight overlap between adjacent class boundaries (e.g., M9.9 and X1.0) may occur. Consequently, this figure should not be used to infer trends across flare classes; Figure~\ref{enh_magnitude_iso_without_jitter} remains the more appropriate figure for such comparisons. The figure shows that there are a significantly greater number of detections associated with X-class flares, compared to M-, C-, and B-class flares.

\begin{figure}[h!]
    \centering
    \includegraphics[width=0.83\textwidth]{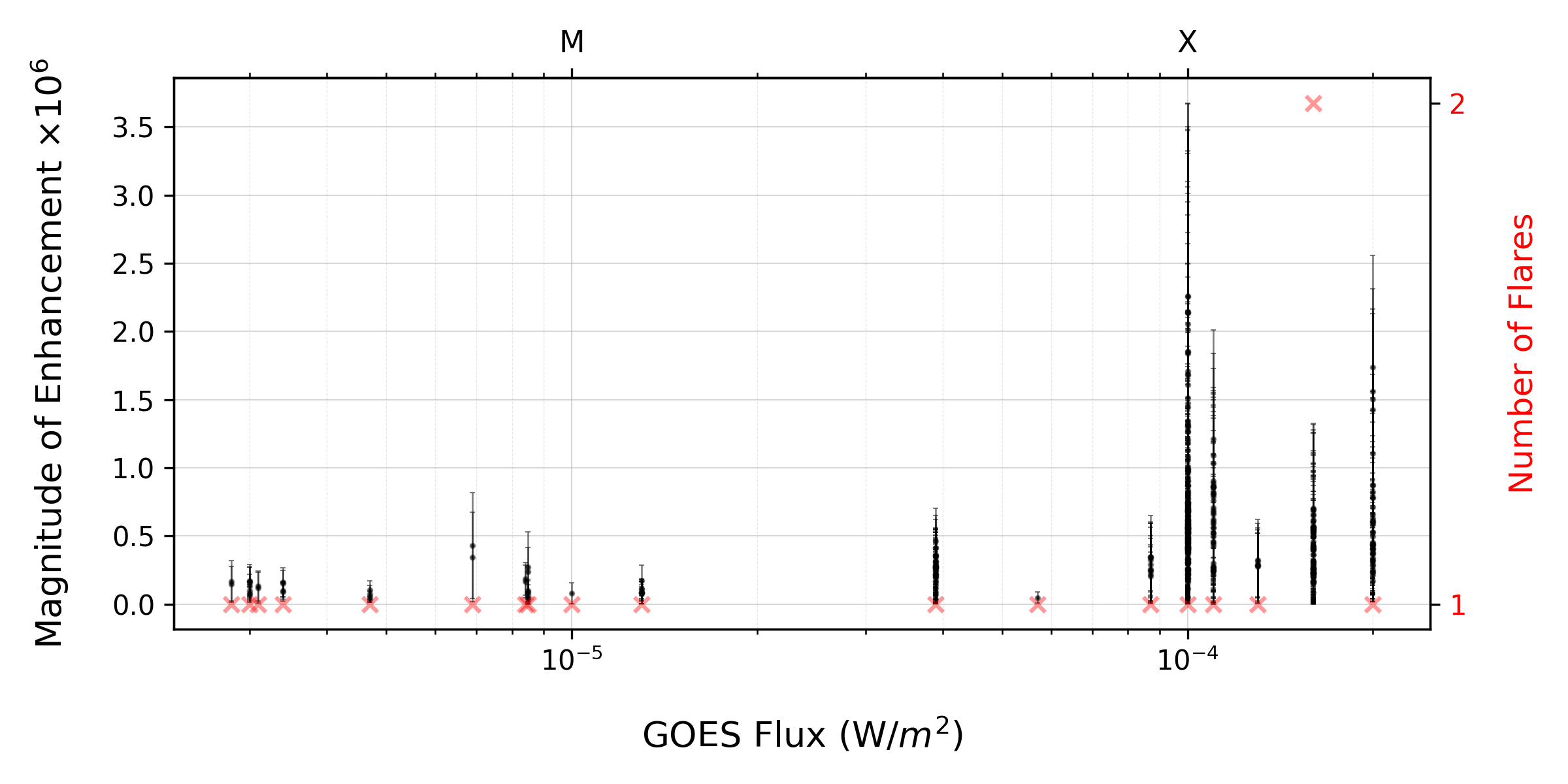}\vspace{-4mm}
    \caption{Magnitude of $\geq 3\sigma$ NUV continuum enhancements as a function of GOES flare flux, detected using the intensity threshold detection pipeline, plotted without jitter. Marker definitions are consistent with those in Fig.~\ref{enh_magnitude_iso_without_jitter}. With the exception of one C-class flare showing a wide prediction interval, the enhancement magnitude generally increases with flare class. A few flares show significantly fewer and weaker detections, which is likely due to limited coverage of the flare ribbons by the IRIS slit.}  
    \label{enh_magnitude_nsigma}
\end{figure}

Figure~\ref{enh_magnitude_nsigma} shows the enhancements detected using the intensity threshold pipeline, which results in fewer overall detections, but with narrower uncertainty intervals. Apart from a single C-class flare with a large uncertainty, the trend of increasing enhancement magnitude with increasing flare strength appears clearer in the results using this pipeline.

\newpage

\newcolumntype{Y}{>{\hspace{0.5em}}X<{\hspace{0.5em}}}
\renewcommand{\arraystretch}{0.2}
\begin{table}[htbp]
\centering
\caption{
Summary of IRIS flares with detected near-ultraviolet (NUV) enhancements above the $3\sigma$ level using IF pipeline. The enhancement time indicates the duration over which enhancements were detected; however, enhancements may not persist throughout the entire interval. In some cases, no corresponding flare was reported by GOES during the listed enhancement interval, which may result in mis-attribution.}
\begin{tabularx}{\textwidth}{@{}c@{\hspace{1em}}Y@{\hspace{1em}}c@{\hspace{1em}}Y@{}}
\toprule
No. & IRIS observation ID & Flare class & Enhancement time (UTC)\\
\midrule
1 & 20140211\_125651\_3860259281 & C8.4 & 13:08:59 -- 13:40:37 \\
\addlinespace
2 & 20140314\_222929\_3860261253 & C3.1 & 00:22:58 -- 00:24:54 \\
\addlinespace
3 & 20140329\_140938\_3860258481 & C1.1 & 16:23:50 -- 16:25:33 \\
\addlinespace
4 & 20140329\_140938\_3860258481 & X1.0 & 17:45:36 -- 17:51:41 \\
\addlinespace
5 & 20140611\_181927\_3863605329 & C3.6 & 18:43:55 -- 18:44:17 \\
\addlinespace
6 & 20140611\_181927\_3863605329 & M3.9 & 20:55:52 -- 21:03:06 \\
\addlinespace
7 & 20140612\_110933\_3863605329 & M1.3 & 18:06:49 -- 18:09:36 \\
\addlinespace
8 & 20140612\_184427\_3863605329 & C8.5 & 00:33:05 -- 00:33:21 \\
\addlinespace
9 & 20140612\_184427\_3863605329 & M1.0 & 21:03:49 \\
\addlinespace
10 & 20140612\_184427\_3863605329 & M3.1 & 22:40:01 -- 22:43:01 \\
\addlinespace
11 & 20140909\_111018\_3860259453 & C3.2 & 12:22:09 -- 12:22:46 \\
\addlinespace
12 & 20140910\_112825\_3860259453 & X1.6 & 17:25:25 -- 17:36:32 \\
\addlinespace
13 & 20141019\_145935\_3860354980 & C4.7 & 17:32:54 \\
\addlinespace
14 & 20141021\_062046\_3860259353 & C2.9 & 06:34:33 -- 06:36:44 \\
\addlinespace
15 & 20141021\_181052\_3860261353 & M8.7 & 01:38:49 -- 01:39:21 \\
\addlinespace
16 & 20141022\_081850\_3860261381 & C3.2 & 11:48:14 \\
\addlinespace
17 & 20141022\_081850\_3860261381 & C5.7 & 17:15:00 \\
\addlinespace
18 & 20141022\_081850\_3860261381 & X1.6 & 14:05:49 -- 14:40:46 \\
\addlinespace
19 & 20141025\_145828\_3880106953 & X1.0 & 17:00:31 -- 17:20:06 \\
\addlinespace
20 & 20141025\_230138\_3864109353 & C3.1 & 01:10:55 -- 01:11:14 \\
\addlinespace
21 & 20141025\_230138\_3864109353 & C4.0 & 05:21:22 -- 05:48:36 \\
\addlinespace
22 & 20141025\_230138\_3864109353 & C8.4 & 23:22:27 -- 23:25:03 \\
\addlinespace
23 & 20141026\_185250\_3864111353 & C4.9 & 05:31:15 -- 05:35:34 \\
\addlinespace
24 & 20141026\_185250\_3864111353 & C9.6 & 06:39:34 -- 06:41:11 \\
\addlinespace
25 & 20141026\_185250\_3864111353 & M1.0 & 02:01:41 \\
\addlinespace
26 & 20141027\_140420\_3860354980 & M1.4 & 17:00:27 -- 17:00:33 \\
\addlinespace
27 & 20141027\_140420\_3860354980 & X2.0 & 14:17:17 -- 14:26:04 \\
\addlinespace
28 & 20141027\_205655\_3864111353 & C5.4 & 21:05:34 -- 21:08:32 \\
\addlinespace
29 & 20141027\_205655\_3864111353 & M3.4 & 02:15:35 -- 02:20:11 \\
\addlinespace
30 & 20150310\_150503\_3860109271 & C2.3 & 22:52:42 -- 22:53:19 \\
\addlinespace
31 & 20150310\_150503\_3860109271 & C3.3 & 23:30:12 \\
\addlinespace
32 & 20150311\_044603\_3860259280 & M2.6 & 07:34:08 \\
\addlinespace
33 & 20150311\_151947\_3860107071 & C1.7 & 16:16:06 \\
\addlinespace
34 & 20150312\_054519\_3860107053 & C4.3 & 08:02:19 \\
\addlinespace
35 & 20150313\_045956\_3860109053 & C6.9 & 07:34:45 -- 07:35:22 \\
\addlinespace
36 & 20150313\_045956\_3860109053 & M1.8 & 05:51:42 -- 05:52:00 \\
\addlinespace
37 & 20150507\_100125\_3860259280 & C4.3 & 10:40:26 -- 10:41:41 \\
\addlinespace
38 & 20150507\_164639\_3860260280 & C2.8 & 16:53:13 \\
\addlinespace
39 & 20150826\_135504\_3860609371 & C5.0 & 14:48:34 \\
\addlinespace
40 & 20150827\_053725\_3860605380 & M2.9 & 05:59:33 \\
\addlinespace
41 & 20150916\_110732\_3680088902 & B9.3 & 13:24:55 \\
\addlinespace
42 & 20220330\_161411\_3660259102 & X1.3 & 17:30:46 -- 17:35:42 \\
\addlinespace
43 & 20220504\_073000\_3884855852 & C3.9 & 10:30:46 -- 10:31:52 \\
\addlinespace
44 & 20220519\_073106\_3680108623 & M1.5 & 10:06:44 \\
\addlinespace
45 & 20220819\_091419\_3660259532 & C3.0 & 10:24:29 \\
\addlinespace
46 & 20220826\_232019\_3660259533 & M4.8 & 02:30:58 \\
\addlinespace
47 & 20220902\_195925\_3660259102 & C3.4 & 20:21:37 -- 20:22:24 \\
\addlinespace
48 & 20220903\_201856\_3660259102 & C2.8 & 21:34:16 \\
\addlinespace
49 & 20230211\_150427\_3660259533 & X1.1 & 15:44:57 -- 15:50:30 \\

\bottomrule
\label{obs_table}
\end{tabularx}
\end{table}
\end{appendix}
\end{document}